%% file: bytecard.tex
\documentclass[sigconf, nonacm]{acmart}

\usepackage{algorithm,algpseudocode}
\usepackage{array,multirow}
\usepackage{xspace}
\usepackage{pifont}
\usepackage{adjustbox}
\usepackage{xcolor}
\usepackage{listings}
\usepackage{colortbl}
\usepackage{pbox}
\usepackage{graphicx}
\usepackage{arydshln}
\usepackage{subfigure}
\usepackage{bm}
\usepackage{amsmath}
\usepackage{makecell}
\usepackage{url}
\usepackage{hyperref}
\usepackage[normalem]{ulem}
\usepackage{float}
\usepackage{setspace}
\usepackage{enumitem}

\usepackage[framemethod=tikz]{mdframed}

\AtBeginDocument{%
  }
    
\renewcommand{\subfigcapskip}{-1em}
\renewcommand{\subfigbottomskip}{-1em}

\newcommand{\CE}{\textsf{CardEst}\xspace}

\newcommand{\FS}{\textsf{ByteCard}\xspace}
\newcommand{\BC}{{ByteHouse}\xspace}
\newcommand{\COUNT}{\textsc{\small COUNT}\xspace}
\newcommand{\COUNTD}{\textsc{\small COUNT-DISTINCT}\xspace}
\newcommand{\LNE}{\textsf{RBX}\xspace}
\newcommand{\IE}{\textsf{Inference Engine}\xspace}
\newcommand{\MS}{\textsf{ModelForge Service}\xspace}
\newcommand{\MPP}{\textsf{Model Preprocessor}\xspace}
\newcommand{\ML}{\textsf{Model Loader}\xspace}
\newcommand{\MV}{\textsf{Model Validator}\xspace}
\newcommand{\MM}{\textsf{Model Monitor}\xspace}
\newcommand{\FJ}{{FactorJoin}\xspace}

\theoremstyle{plain}
\newtheorem*{example*}{Example}

\definecolor{mygrey}{RGB}{230,230,240}

\definecolor{commentgreen}{RGB}{0,128,0}
\definecolor{keywordblue}{RGB}{0,0,255}
\definecolor{stringred}{RGB}{255,0,0}
\definecolor{backgroundgray}{RGB}{240,240,240}
\definecolor{numberorange}{RGB}{255,165,0}
\definecolor{preprocessorpurple}{RGB}{153,50,204}
\definecolor{mykeywords}{RGB}{153, 0, 0}

\if@ACM@journal\else
\acmConference[Conference’24]{ACM Conference}{June 2024}{Santiago, Chile}%
\fi

\lstdefinelanguage{mylanguage}{
	basicstyle=\ttfamily,
    showstringspaces=false
    sensitive=false,
    keywords=[0]{class, virtual, double, void},
    keywords=[1]{CardinalityEstimator},
    comment = [l]{//}
}

\setcopyright{acmcopyright}
\copyrightyear{2024}
\acmYear{2024}
\acmDOI{XXXXXXX.XXXXXXX}

\acmPrice{15.00}
\acmISBN{978-1-4503-XXXX-X/18/06}

\begin{document}


\title{\FS:~Enhancing ByteDance's Data Warehouse with Learned Cardinality Estimation}


\author{Yuxing Han$^*$}
\affiliation{%
  \institution{Data Platform, ByteDance}
  \city{Shanghai}
  \country{China}
  }
\email{hanyuxing@bytedance.com}

\author{Haoyu Wang}
\author{Lixiang Chen}
\affiliation{%
  \institution{Data Platform, ByteDance}
  \city{Shanghai}
  \country{China}
}
\affiliation{%
  \institution{East China Normal University}
  \city{Shanghai}
  \country{China}
  }
\email{wanghaoyu.0428@bytedance.com}
\email{chenlixiang.3608@bytedance.com}

\author{Yifeng Dong}
\author{Xing Chen}
\affiliation{%
  \institution{Data Platform, ByteDance}
  \city{Beijing}
  \country{China}
  }
\email{dongyifeng@bytedance.com}
\email{chenxing.xc@bytedance.com}

\author{Benquan Yu}
\affiliation{%
  \institution{Data Platform, ByteDance}
  \city{Shanghai}
  \country{China}
  }
\email{james.yu@bytedance.com}

\author{Chengcheng Yang$^*$}
\affiliation{%
  \institution{East China Normal University}
  \city{Shanghai}
  \country{China}
  }
\email{ccyang@dase.ecnu.edu.cn}

\author{Weining Qian}
\affiliation{%
  \institution{East China Normal University}
  \city{Shanghai}
  \country{China}
  }
\email{wnqian@dase.ecnu.edu.cn}


\input{sections/abstract}

\maketitle

\begingroup
\renewcommand\thefootnote{}\footnote{\noindent
	$*$ Corresponding authors
}
\addtocounter{footnote}{-1}
\endgroup

\renewcommand{\shortauthors}{Yuxing Han et al.}

\input{sections/introduction.tex}

\input{sections/related.tex}

\input{sections/choice.tex}

\input{sections/framework.tex}

\input{sections/adapt.tex}

\input{sections/applications}

\input{sections/experiments.tex}

\input{sections/lessons.tex}

\input{sections/conclusions.tex}


\bibliographystyle{ACM-Reference-Format}
\bibliography{ref}

\end{document}

%% file: sections/abstract.tex
\begin{abstract}
Cardinality estimation is a critical component and a longstanding challenge in modern data warehouses.
\BC, ByteDance's cloud-native engine for extensive data analysis in exabyte-scale environments, serves numerous internal decision-making business scenarios. 
With the increasing demand for \BC,  cardinality estimation becomes the bottleneck for efficiently processing queries. 
Specifically, the existing query optimizer of \BC uses the traditional Selinger-like cardinality estimator, which can produce substantial estimation errors, resulting in suboptimal query plans.

To improve cardinality estimation accuracy while maintaining a practical inference overhead, we develop a framework \FS that enables efficient training and integration of learned cardinality estimators.
Furthermore, \FS adapts recent advances in cardinality estimation to build models that can balance accuracy and practicality (e.g., inference latency, model size, training overhead).
We observe significant query processing speed-up in \BC after replacing the existing cardinality estimator with \FS for several optimization scenarios. Evaluations on real-world datasets show the integration of \FS leads to an improvement of up to 30\% in the 99th quantile of latency.
At last, we share our valuable experience in engineering advanced cardinality estimators.
This experience can help \BC integrate more learning-based solutions on the critical query execution path in the future.
\end{abstract}

%% file: sections/introduction.tex
\section{Introduction}

\BC, ByteDance's internal data warehouse, is crucial for handling analytics at an exabyte scale, underpinning various business decisions through applications that include risk management and strategic marketing. Building on the collective research insights and engineering endeavors of predecessors~\cite{stonebraker2018c,armenatzoglou2022amazon,lamb2012vertica,sun2023presto,dageville2016snowflake,HuangLCFMXSTZHW20}, \BC has demonstrated robust performance across a range of business workloads. 
In its consistent pursuit of excellence, \BC continually seeks to evolve and improve, particularly in addressing the cardinality estimation (\CE) challenge---a critical aspect of query optimization that has received extensive attention from academia and industry. Cardinality estimation aims to estimate query operator results size without actual execution, consisting of \COUNT and \COUNTD (NDV) estimation. Accurate estimation approaches are important for enhancing query plans' quality, which is one of \BC's most notable performance bottlenecks.

\begin{table*}[ht]
\caption{Estimation Errors of Traditional \CE Approaches in \BC}
\vspace{-1.5em}
\begin{spacing}{1} 
\scalebox{0.9} {
\begin{tabular}{p{\dimexpr0.15\textwidth-1\tabcolsep}|p{\dimexpr0.1\textwidth-2\tabcolsep}p{\dimexpr0.1\textwidth-2\tabcolsep}p{\dimexpr0.1\textwidth-2\tabcolsep}p{\dimexpr0.1\textwidth-2\tabcolsep}p{\dimexpr0.1\textwidth-2\tabcolsep}p{\dimexpr0.1\textwidth-2\tabcolsep}p{\dimexpr0.1\textwidth-2\tabcolsep}p{\dimexpr0.1\textwidth-2\tabcolsep}p{\dimexpr0.1\textwidth-2\tabcolsep}}
\hline
\rowcolor{mygrey} 
\cellcolor{mygrey}                                  & \multicolumn{3}{c}{\cellcolor{mygrey}\textbf{IMDB}}  & \multicolumn{3}{c}{\cellcolor{mygrey}\textbf{STATS}} & \multicolumn{3}{c}{\cellcolor{mygrey}\textbf{AEOLUS}} \\ \cline{2-10} 
\rowcolor{mygrey} 
\multirow{-2}{*}{\cellcolor{mygrey}\textbf{CardEst}} & \textbf{50\%} & \textbf{90\%} & \textbf{99\%}                & \textbf{50\%} & \textbf{90\%} & \textbf{99\%}                & \textbf{50\%}       & \textbf{90\%}      & \textbf{99\%}      \\ \hline
COUNT Est.                                               & $3.06$          & $1145$          & \multicolumn{1}{l|}{$1\cdot10^6$} & $493$          & $3\cdot10^4$      & \multicolumn{1}{l|}{$3\cdot10^7$} & $7.45$                & $3 \cdot 10^6$               & $8\cdot10^6$            \\
\hline
NDV Est.                                               & $15$            & $984$           & \multicolumn{1}{l|}{$3\cdot10^4$} & $134$           & $1\cdot10^4$      & \multicolumn{1}{l|}{$6\cdot10^4$} & $598$                 & $4912$               & $2\cdot10^4$  \\
 \hline
\end{tabular}}
\end{spacing}
\vspace{-1.2em}
\label{tab: qerror-report}

\end{table*}

In its initial stages, \BC employed traditional \CE approaches like other modern data warehouses. However, the inherent data skewness in real-world datasets, coupled with the simplified assumptions of these approaches, 
hindering \BC from achieving reliable estimates. This problem is further exacerbated when encountering large volumes of customer data and rapid data updates. Traditional sketch-based approaches~\cite{selinger1979access,flajolet2007hyperloglog} 
often require full data scans, creating considerable pressure on \BC's storage layer. Meanwhile, the sample-based approaches face inherent
challenges in balancing accuracy with the sampling rate.

The evaluation report in Table~\ref{tab: qerror-report} shows the inadequacy of traditional \CE approaches in handling current analytical workloads.
This report evaluates various quantiles of the Q-Error, a widely used metric in evaluating \CE approach ~\cite{leis2015good,leis2018query}, known for its theoretical lower bound of $1$.
Specifically, the datasets under consideration include two benchmarks, IMDB~\cite{MSCN} and STATS~\cite{cardestbench}, as well as AEOLUS, an internal business workload from \BC that comprises $200$ complex queries from customers.
The evaluation results indicate that for both \COUNT and \COUNTD estimation, the errors of traditional approaches deviate far from the theoretical optimal lower bound across various quantiles, often by several orders of magnitude. This discrepancy highlights a significant potential for improving the estimation methodology.

Recently, learning-based \CE approaches~\cite{hilprecht2019deepdb,yang2020neurocard,zhu2020flat,negi2023robust,wu2023factorjoin,Cohen2019CardinalityEI,ndvwu2021,MSCN,wu2020bayescard,yang2019deep,liu2021fauce,dutt2019selectivity,wu2021unified} 
have drawn much attention due to their superior accuracy~\cite{cardestbench,sun2021learned,thirumuruganathan2022prediction}. The prosperity of these \CE research work naturally raises a question: Could we replace traditional approaches in \BC with learning-based ones to get more accurate cardinality estimates, thereby enhancing query optimization? After a deep survey of the learned approaches, we identify three challenges of integrating learning-based estimators into the \BC's existing architecture:
(1) \textit{How to discern the appropriate estimation models that balance accuracy and practicability?} 
Although the existing studies have proposed numerous learned estimators, most focus on improving accuracy.
The goal of \FS is to achieve accurate estimation and improve \BC's overall query performance in a resource-efficient manner.
(2) \textit{How to manage the training process and integrate the inference algorithms of these models in the query processing?} 
Due to the necessity of not disrupting customers' active online queries, training directly on large-scale data volumes stored in \BC is impractical. Furthermore,
deploying existing inference algorithms within multi-threaded execution environments poses another challenge.
(3) \textit{How to utilize the models' estimates to enable enhanced query optimization for \BC?}
Identifying optimization scenarios in \BC that suffer from poor estimation approaches and applying accurate estimates from \FS for enhanced query optimization presents a non-trivial challenge.
 
In this paper, we present \FS, an enhanced \CE framework designed to integrate learning-based approaches into \BC seamlessly.
To strike a balance between accuracy and practicality, we first select models by carefully evaluating inference latency, model size, and the training/updating overhead for optimization scenarios in \BC.
Then, \FS introduces an abstraction engine to ease the integration of inference algorithms for different models.
Besides, the engine could also help identify immutable data structures, which would further avoid data races and enable high-concurrency inference executions in the multi-threaded environment.
In addition, \FS also proposes a dedicated service for isolated training to ensure no disruption on the online queries. Moreover, \FS employs auxiliary modules for model loading and accuracy monitoring to sustain its effectiveness. 
As a result, applying \FS's estimation in \BC's optimization scenarios has significantly accelerated query processing.
For now, \FS is responsible for millions of cardinality estimations within \BC's online clusters, providing substantial benefits to various analytical workloads.
As a pioneer in integrating learning-based approaches into a production-scale data warehouse, \FS demonstrates the significant potential of machine learning to enhance query optimization in large-scale systems.

In summary, our main contributions are listed as follows:
\begin{itemize}

	\item We make careful model choices for learned cardinality estimators by evaluating factors such as inference latency and training overhead for effective integration into \BC.

	\item We introduce \FS, a framework designed to integrate learning-based cardinality estimators in ByteDance's internal data warehouse, \BC. The framework features an inference abstraction engine and a dedicated training service, which facilitate the integration of chosen models.

	\item The accurate cardinality estimates provided by \FS are applied in several optimization scenarios and have demonstrated their effectiveness in enhancing \BC's query performance in subsequent evaluations.
	
	\item We share lessons derived from the design, development, and deployment experience of \FS, along with our future efforts to integrate more learning-based approaches to further enhance \BC's query optimization. 

\end{itemize}

%% file: sections/related.tex
\section{Related Work}

\noindent \underline{\textbf{Learning-Based \CE Methods:}}
The research community has proposed a diverse set of learned models~\cite{zhu2020flat,wu2020bayescard,yang2020neurocard,hilprecht2019deepdb,wu2023factorjoin,dutt2019selectivity,liu2021fauce,ndvwu2021,Cohen2019CardinalityEI} for both \COUNT and \COUNTD estimation. Based on recent studies~\cite{cardestbench,zhu2020flat}, they can be broadly classified into query-driven and data-driven methods.
The query-driven methods~\cite{MSCN, liu2021fauce, suncostmodels2019, dutt2019selectivity} aim to map each featurized query to its \COUNT or \COUNTD cardinality, utilizing advanced models like gradient boosted trees \cite{dutt2019selectivity} and DNNs \cite{Cohen2019CardinalityEI, MSCN}. In contrast, the data-driven methods~\cite{zhu2020flat, hilprecht2019deepdb, yang2020neurocard, wang2021face, wu2020bayescard} treat table tuples as samples from a joint distribution, applying ML-based models such as deep auto-regression \cite{yang2019deep, yang2020neurocard}, Bayesian Networks \cite{wu2020bayescard}, and Sum-Product Networks \cite{hilprecht2019deepdb, zhu2020flat}. Another model, \LNE(named after the initials of its first three authors' surnames)~\cite{ndvwu2021}, offers a workload-independent approach for NDV estimation. Selecting the most suitable models for \FS's requirements is a crucial priority.

\noindent \underline{\textbf{ML-enhanced Components of Databases:}}
Recent efforts in both academia and industry have focused on harnessing machine learning to boost database system performance. 
SageDB \cite{ding2022sagedb} is a data analytics prototype that utilizes learned components to self-tune for optimal performance across various datasets and queries, focusing on machine learning techniques like partial materialized views and global optimization algorithms.
Bourbon \cite{dai2020wisckey}, a learned index for LSM trees, uses piecewise linear regression for key distribution learning to enhance the lookup efficiency. It also offers guidance for integrating learned indexes into LSM trees tailored to specific levels and workloads.
Xindex \cite{tang2020xindex} is a concurrent ordered index optimized for quick queries, employing a hierarchical structure that adapts to real-time workloads, outperforming traditional index structures in efficiency.
OpenGauss \cite{opengauss} integrates machine learning for various self-management tasks, including query rewriting, cost estimation, and plan generation.
Auto-WLM \cite{saxena2023auto} is a machine learning-based workload management system employed in Amazon Redshift \cite{armenatzoglou2022amazon}, which dynamically schedules workloads and adjusts to changes, using local query performance models to enhance overall cluster performance.
We posit that \FS represents the first instance of integrating learning-based cardinality estimation models into an industrial data warehouse system.

%% file: sections/choice.tex
\section{Background and model choices}
\label{sec: choice}
This section begins with an overview of \BC, followed by an analysis of several optimization scenarios hindered by the poor cardinality estimation approaches. At last, we thoroughly discuss our selection process for \FS's learned \CE models.

\subsection{Overview of \BC}
\begin{figure}[t]
	\centering
	\includegraphics[width=9cm]{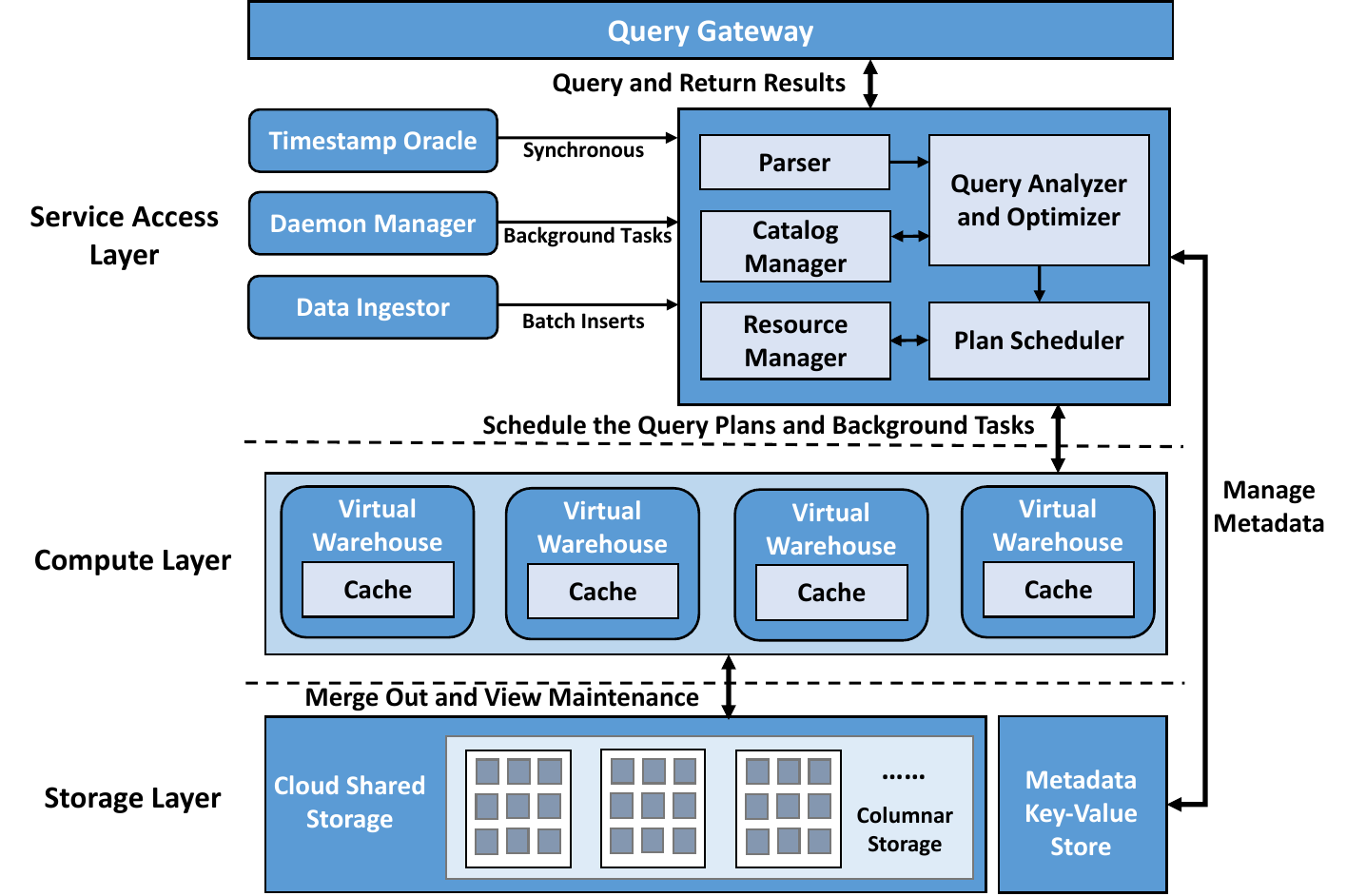}
        \vspace{-2em}
	\caption{The Architecture of \BC}
	\label{fig: cnch}
\end{figure}

The high-level architecture of \BC is shown in \ref{fig: cnch}.
\BC employs an architecture that separates storage from computation, consisting of three discrete layers: service, computing, and storage. This modular structure facilitates a clear division of responsibilities, enhancing scalability and efficiency within the system.
The service layer parses queries, optimizes execution plans, and dispatches tasks to computing nodes. The computing layer dynamically allocates resources and executes query operators. The storage layer consists of a distributed key-value store for managing metadata and a distributed file system for storing business data.
Core components include the Resource Manager, which orchestrates the allocation of computing resources; the Time Oracle, which ensures synchronized computation operations; the Data Ingestor, which manages data flow from multiple sources; and the Daemon Manager, which manages the lifecycle of background tasks.
In addition, \BC utilizes other widely adopted techniques in modern data warehouses,  such as 
columnar storage, and vectorized execution.

\subsubsection{Optimization Scenarios Hindered by Inaccurate Cardinality Estimation} 
Many optimization scenarios have been developed across \BC's three layers to ensure good performance for its users.
These scenarios include sideways information passing~\cite{ives2008sideways}, magic set rewriting~\cite{seshadri1996cost}, and late materialization~\cite{abadi2006materialization,shrinivas2013materialization}.
However, their effectiveness is limited by the inaccuracies of traditional \CE approaches. 
We explore two examples in the following discussions.

\begin{table*}[htb]
\caption{Estimation Errors of Learned \CE Approaches in \FS}
\vspace{-1em}
\begin{spacing}{1} 
\scalebox{0.9} {
\begin{tabular}{p{\dimexpr0.15\textwidth-1\tabcolsep}|p{\dimexpr0.1\textwidth-2\tabcolsep}p{\dimexpr0.1\textwidth-2\tabcolsep}p{\dimexpr0.1\textwidth-2\tabcolsep}p{\dimexpr0.1\textwidth-2\tabcolsep}p{\dimexpr0.1\textwidth-2\tabcolsep}p{\dimexpr0.1\textwidth-2\tabcolsep}p{\dimexpr0.1\textwidth-2\tabcolsep}p{\dimexpr0.1\textwidth-2\tabcolsep}p{\dimexpr0.1\textwidth-2\tabcolsep}}
\hline
\rowcolor{mygrey} 
\cellcolor{mygrey}                                  & \multicolumn{3}{c}{\cellcolor{mygrey}\textbf{IMDB}}  & \multicolumn{3}{c}{\cellcolor{mygrey}\textbf{STATS}} & \multicolumn{3}{c}{\cellcolor{mygrey}\textbf{AEOLUS}} \\ \cline{2-10} 
\rowcolor{mygrey} 
\multirow{-2}{*}{\cellcolor{mygrey}\textbf{CardEst}} & \textbf{50\%} & \textbf{90\%} & \textbf{99\%}                & \textbf{50\%} & \textbf{90\%} & \textbf{99\%}                & \textbf{50\%}       & \textbf{90\%}      & \textbf{99\%}      \\ \hline
COUNT Est.                                                  & $1.14$          & $4.82$          & \multicolumn{1}{l|}{$425$}      & $1.47$          & $8.03$          & \multicolumn{1}{l|}{$4026$}     & $1.3$                 & $3.57$               & $7491$                \\ 
\hline
NDV Est.                                                  & $3.67$          & $191$           & \multicolumn{1}{l|}{$392$}      & $2.93$          & $362$           & \multicolumn{1}{l|}{$517$}      & $3.8$                 & $133$                & $934$                 \\
 \hline
\end{tabular}}
\end{spacing}
\vspace{-1em}
\label{tab: qerror-report-learned}

\end{table*}

\noindent \underline{\textbf{Materialization Strategies:}}
During query processing in column-based systems like \BC, materialization involves converting data from a columnar format to a row-based tuple. 
Regarding materialization strategies, \BC adopts the method of early materialization, where tuples are generated early in the query plan, owing to its simplicity and common use~\cite{sun2023presto,shrinivas2013materialization}.
Initially, \BC adopted a one-stage reader approach for early materialization, which involves scanning and processing queried columns in one pass, applying all necessary predicates simultaneously.
The method is effective for non-selective predicates as it reduces per-tuple processing by handling entire column blocks. 
However, the current method becomes less efficient with highly selective predicates, where it constructs many unnecessary tuples, even though only a small subset of tuples require further processing. 
Therefore, accurate selectivity estimation becomes crucial in optimizing the materialization strategies.
Traditional \CE approaches often struggle to provide accurate cardinality/selectivity due to their limited capability in identifying cross-column correlations compared to learned ones, especially with small sample sizes.

\noindent \underline{\textbf{Aggregation Processing:}}
 The common practice of aggregation processing in data warehouses usually involves employing an in-memory hash table to record the distinct values of aggregation keys.
Effectively managing this hash table often involves addressing the issue of handling the increasing values of aggregation keys. 
The \textit{resizing} operation, which requires allocating larger memory blocks and rehashing entries, is resource-intensive due to the consumption of significant CPU and memory. We observe that frequent early-stage \textit{resizing}s incur notable overhead, adversely affecting \BC's performance of aggregation processing. 
To address this issue, the current strategy employed by \BC is to cache the size of hash tables of previous queries. 
However, this method is effective only for identical repeat queries and quickly loses its effectiveness due to data updates. 
An alternative method is to reduce the frequency of \textit{resizing} by accurately estimating the initial size of the required hash table. Therefore, a highly accurate NDV estimator is pivotal in minimizing the frequent \textit{resizing}s operations.

\subsection{\CE Model Choice}
Selecting the most suitable estimation models from the various options available is a crucial challenge for \FS, considering their practical integration into \BC's existing architecture. The ideal models for the integration must fulfill the following criteria:  1) They should offer higher accuracy in cardinality estimates than traditional approaches in most cases;  2) Given the limited resources for cardinality estimation in query optimization, the training process should be resource-efficient;  3) The inference algorithm needs to be efficient enough to avoid heavily influencing the query execution time. After careful analysis and evaluation, \FS selects three learned \CE models, each optimized for specific optimization scenarios in \BC.

\subsubsection{Choice for \COUNT \CE Models}
Query-driven models like MSCN~\cite{MSCN}, which necessitate extensive query logs and computation of true cardinality for training, are resource-intensive.
Besides, these models may diminish effectiveness with data changes as they are specific to certain workloads.
Therefore, query-driven methods are considered impractical for \FS's requirements.
Alternatively, data-driven methods focus on learning data distributions through unsupervised ML models. 
For example, BayesCard \cite{wu2020bayescard} offers a solution with its tree-structured Bayesian Networks (BNs), addressing single-table \COUNT estimations. These networks stand out for their advantages of high accuracy, efficient training, small model size, quick inference, and adaptability to data changes \cite{cardestbench}.

However, to handle join-size estimation, most of the data-driven methods~\cite{wu2020bayescard, zhu2020flat, hilprecht2019deepdb, yang2020neurocard, wang2021face}, including BayesCard, hold the design philosophy to understand the joint distribution of the joined tables, imposing a non-trivial overhead for model training. 
Besides, these methods usually employ the denormalizing strategy, which will add extra columns to facilitate later inference. 
The number of extra columns will expand rapidly as the number of join relationships increases. 
This is not affordable inside \BC as the system often needs to handle numerous complex join relationships between different tables. 
Therefore, we decide to adopt a recent approach \FJ~\cite{wu2023factorjoin} for join queries. 
It naturally supports using the Bayesian Networks as simple-table cardinality estimation and requires almost no additional training overhead.
Specifically, in the offline training phase, \FJ 
creates specialized buckets on the join key values (i.e., join-buckets), and builds Bayesian Networks to understand the correlations among filter columns and join keys within a single table.
In the online inference phase, \FJ first dynamically
constructs a factor graph \cite{loeliger2004introduction}, which is derived from the join relationships specified in the query. Then, it utilizes the related simple-table Bayesian Networks and applies inference on the graph with join-buckets to estimate the cardinality bounds accurately.
\FJ is feasible for integration in \FS due to its efficient training process and proven advantage in estimation accuracy and inference speed over alternative methods~\cite{wu2023factorjoin}.

\subsubsection{Choice for \COUNTD \CE Models}

The traditional \COUNTD (NDV) \CE approaches can be classified into sketch-based and sampling-based categories. However, both of them face challenges when dealing with small samples.
The commonly used sketch-based estimator HyperLogLog (HLL)~\cite{flajolet2007hyperloglog,heule2013hyperloglog} has no theoretical guarantees for sampled data and usually requires a full dataset scan for accurate estimation.
Moreover, frequent data updates reduce the effectiveness of old sketches.
Meanwhile, the sample-based estimators~\cite{charikar2000towards,chao1992estimating} often rely on specific heuristics or data assumptions, which might not generally apply to diverse datasets. Their robustness is compromised as the foundational assumptions are prone to breakdown.

The learning-based NDV estimator proposed in~\cite{Cohen2019CardinalityEI} adopts a supervised learning framework, requiring the collection of true NDV from a significant number of online queries for each workload.
Alternatively, \LNE~\cite{ndvwu2021} adopts a one-model-fits-all approach that treats NDV as a standard data property akin to standard deviation in statistics, and aims to derive a ``closed''-formula of NDV calculation. This approach opts for a neural network to learn this formula with the belief it can approximate any continuous function.
This estimator exhibits robust performance across a spectrum of workloads and maintains effectiveness across different sampling rates, fulfilling the accuracy criterion.
The workload-independent nature guarantees one training process can serve a wide range of workloads, aligning with the resource-efficiency criterion.
Moreover, the neural network designed by \LNE has an acceptable number of network layers, enabling \FS to conduct efficient inference within \BC's query processing.
Therefore, \FS chooses \LNE as its learning-based NDV estimator.

\begin{table*}[htb]
\caption{The Training Time and Model Size between Different \CE Models}
\vspace{-1em}
\begin{spacing}{1.0} 
\scalebox{0.9}{
\begin{tabular}{
>{\raggedright\arraybackslash}m{\dimexpr0.15\textwidth+0\tabcolsep}|
>{\centering\arraybackslash}m{\dimexpr0.07\textwidth-2\tabcolsep}
>{\centering\arraybackslash}m{\dimexpr0.07\textwidth-2\tabcolsep}
>{\centering\arraybackslash}m{\dimexpr0.07\textwidth-1\tabcolsep}
>{\centering\arraybackslash}m{\dimexpr0.07\textwidth-2\tabcolsep}
>{\centering\arraybackslash}m{\dimexpr0.07\textwidth-2\tabcolsep}
>{\centering\arraybackslash}m{\dimexpr0.07\textwidth-1\tabcolsep}
>{\centering\arraybackslash}m{\dimexpr0.07\textwidth-2\tabcolsep}
>{\centering\arraybackslash}m{\dimexpr0.07\textwidth-2\tabcolsep}
>{\centering\arraybackslash}m{\dimexpr0.07\textwidth-1\tabcolsep}
>{\centering\arraybackslash}m{\dimexpr0.07\textwidth-2\tabcolsep}
>{\centering\arraybackslash}m{\dimexpr0.07\textwidth-2\tabcolsep}
>{\centering\arraybackslash}m{\dimexpr0.07\textwidth-0\tabcolsep}}
\hline
\rowcolor{mygrey} 
\cellcolor{mygrey}                               & \multicolumn{3}{c}{\cellcolor{mygrey}\textbf{MSCN}}  & \multicolumn{3}{c}{\cellcolor{mygrey}\textbf{DeepDB}} & \multicolumn{3}{c}{\cellcolor{mygrey}\textbf{BayesCard}} & \multicolumn{3}{c}{\cellcolor{mygrey}\textbf{\FJ}} \\  \cline{2-13} 
\rowcolor{mygrey} 
\multirow{-2}{*}{\cellcolor{mygrey}\textbf{Measure}} & \textbf{IMDB} & \textbf{STATS} & \textbf{AEOLUS}                & \textbf{IMDB} & \textbf{STATS} & \textbf{AEOLUS}                & \textbf{IMDB}       & \textbf{STATS}      & \textbf{AEOLUS}      & \textbf{IMDB}       & \textbf{STATS}      & \textbf{AEOLUS}      \\ \hline
Training Time (Min)                                              & $41$          & $34$          & \multicolumn{1}{c|}{$45$} & $78$          & $113$      & \multicolumn{1}{c|}{$145$} & $33$                & $27$               & \multicolumn{1}{c|}{$31$}            & $7$       & $13$      & $11$      \\ \cline{1-13}
Model Size (MB)                                                 & $3.9$          & $2.8$          & \multicolumn{1}{c|}{$7.3$}      & $43$          & $162$          & \multicolumn{1}{c|}{$201$}     & $2.4$                 & $6.1$               & \multicolumn{1}{c|}{$6.5$}                & $4.3$       & $2.3$      & $3.2$      \\ \hline
\end{tabular}
}
\end{spacing}
\vspace{-0.7em}
\label{tab: training-report}
\end{table*}

\subsubsection{Evaluation \& Summary for Model Choices}

To ascertain the effectiveness of our model choices for \FS, we translate \BC's cardinality estimation into SQL queries when processing the workloads from the IMDB,  STATS, and AEOLUS datasets. Next, we train different models on the three datasets offline. 
The performance of these models is evaluated by comparing their Q-Error results, as shown in Table~\ref{tab: qerror-report-learned}, against those of traditional approaches shown in Table~\ref{tab: qerror-report}.
This comparison showcases the effectiveness of selected models, especially notable at the 99\% quantile, where learning-based approaches demonstrate significant improvements.

Then, we investigate the training time and model size across different datasets for different estimation models. 
Given that \LNE follows a one-model-fits-all approach, its training algorithm is not evaluated here.
We select MSCN as the representative query-driven model alongside three data-driven models: DeepDB, BayesCard, and \FJ. 
The training configurations for all models follow the default specifications. Specifically, for \FJ's bucket strategy, we opt for equi-height buckets with a total count of $200$.
From the results in Table~\ref{tab: training-report}, we can see that the training time of MSCN consistently exceeds that of other models across various datasets. Note that this time does not include the computation time of true cardinalities as training objectives. This observation underscores the impracticality of query-driven models for integration in \FS.
Among data-driven models, DeepDB~\cite{hilprecht2019deepdb} and BayesCard~\cite{wu2020bayescard} exhibit longer training times and larger model sizes, which is attributed to their denormalization strategy for join-size estimation. In contrast, \FJ effectively reduces training overhead and model size while preserving high-accuracy estimations, because it
leverages simple-table models and captures join-key distributions using join-buckets and factor graphs.

\noindent \underline{\textbf{Summary}}: 
\FS employs a lightweight Bayesian Network per table for estimating single-table \COUNT cardinalities and combines these models through \FJ to accurately estimate join sizes. 
For NDV estimation, \FS leverages the workload-independent \LNE approach, where one model from the offline training process is adequate for the majority of estimation scenarios.

%% file: sections/framework.tex
\section{System Architecture of \FS}

In this section, we first outline the design principles of \FS. Then, we give a detailed description of \FS's each module. 

\subsection{Design Principles}

\begin{figure}[tb]
	\centering
	\includegraphics[width=7.9cm]{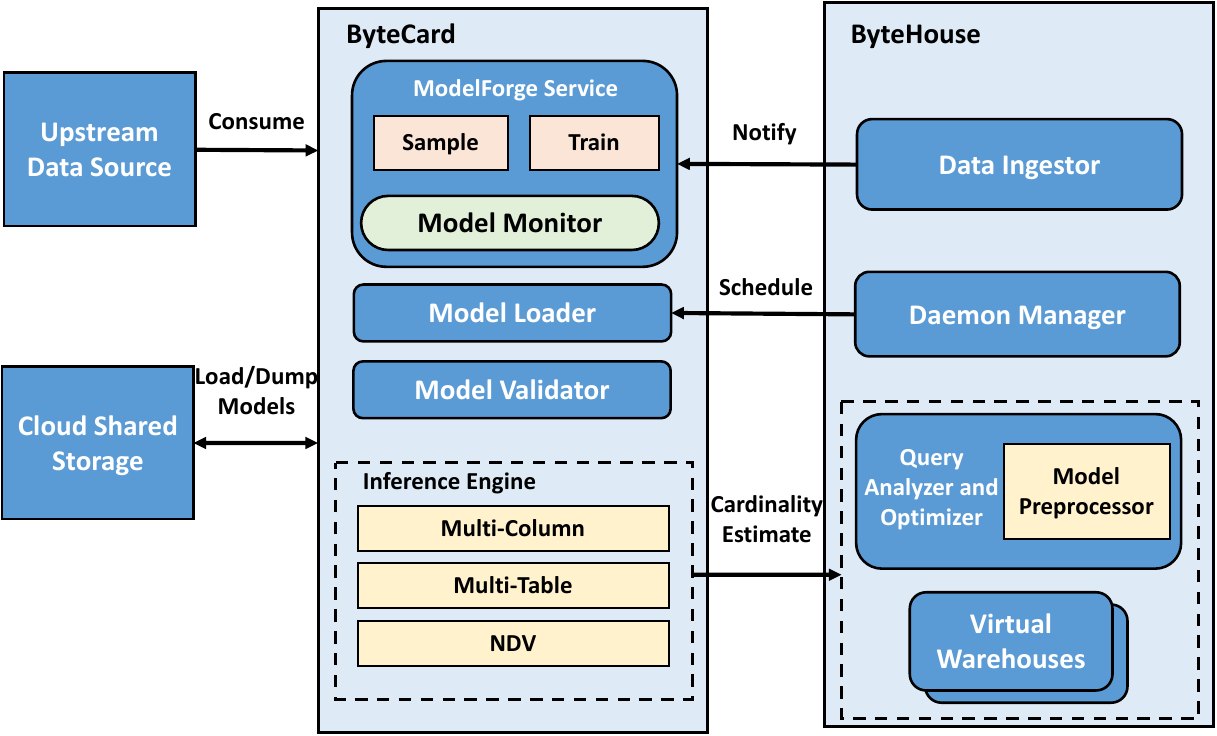}
	\caption{The Architecture of \FS}
	\label{fig: bytecard}
\end{figure}

When developing \FS to optimize \BC's query processing, we emphasize its practicability, which requires reducing computational overhead and ensuring efficient use of learned \CE models. Moreover, we preserve \FS's flexibility to facilitate the integration of new \CE models and broaden the framework to include more learning-based query optimization techniques.
The architecture of \FS is shown in Figure~\ref{fig: bytecard}.
To fulfill the goals described above, \FS introduces two core modules, i.e., a high-level program abstraction known as \IE and a standalone service called \MS, alongside auxiliary modules, to keep \FS's efficiency and effectiveness while safeguarding \BC's stability. 
   
The \IE is the central hub for deploying inference algorithms of \CE models. It aims to simplify the integration process for inference algorithms in multi-threaded query processing environments. 
The \MS is designed to focus on the iterative integration for training and the models' management, ensuring their accuracy.
In addition to the core modules, auxiliary modules are developed, including the \ML and \MM. The \ML is responsible for efficiently loading and updating models across the large-scale cluster. Meanwhile, the \MM takes care of the model quality, triggering models' fine-tuning if necessary. 
The auxiliary modules collectively guarantee the efficient functioning of the \IE and \MS, which further provide strong support for the effectiveness of the \FS framework and the stability of \BC.

\subsection{Inference Engine} \label{sec: inference}
As shown in Figure~\ref{fig: infer_engine}, the \IE provides a high-level abstraction for communication with \FS's other modules and the integration with \BC's query processing. With this engine, \BC can enhance its query optimization by leveraging the accurate estimations provided by the advanced learning-based cardinality estimation models.

\subsubsection{Interact with other modules} \label{sec: type_mapping}

As each selected model in \FS has different structures and serialization methods, the \textsf{loadModel} interface is crafted to encapsulate the deserialization process for different models.
This interface is usually invoked by \ML, a background process responsible for loading \CE models from the cloud-based storage.
From the perspective of ByteHouse’s Daemon Manager, the \ML operates similarly to other background tasks, such as the LSM-tree's compaction task~\cite{lsmanalysis,chen2023workloadaware} in \BC's storage layer. Then, the Daemon Manager assigns resources to these loading tasks similarly, except for the strategy of task triggering.  In contrast to the complex strategies utilized for LSM-tree's compaction, our approach employs a timestamp-based approach for loading the up-to-date model. 
Consequently, \FS guarantees only models with 
the most recent timestamp are considered for loading and updating.
In the current configuration, models are scheduled for loading at a default interval of one hour unless \MM detects that the performance of models is decreased due to the shift of data distribution.

Upon loading a model into memory, the \MV employs its \textsf{validate} interface to evaluate the model's validity. 
This step is crucial for preventing potential crashes during actual inference (i.e., executing the \textsf{estimate} interface) in \BC's query processing.
The validation process involves two primary checks: the \textit{size checker} and the \textit{health detector}.
The \textit{size checker} regulates the size of individual models and the total size of all the loaded models to prevent excessive memory usage in \BC. 
To avoid the case in which one table's model occupies too much memory, \FS will refuse to load a model if its size is too large.
Moreover, when the cumulative size exceeds a predetermined threshold, \FS employs a Least Recently Used (LRU) strategy to prioritize and retain the most frequently used models.
The \textit{health detector} is responsible for maintaining the models' healthy state. For example, in the case of Bayesian Networks, the detector employs a cyclic detection method to verify the structural legitimacy of the model, ensuring its structure conforms to a directed acyclic graph (DAG).
  
After the model is validated successfully, the subsequent step involves employing \textsf{initContext} to establish the programming context for inference algorithms, thereby preparing the model for estimation in the query processing.
This interface enables the initialization of immutable data structures extracted from the inference algorithms, ensuring these structures remain read-only within each query thread. This approach allows the algorithms to be executed lock-free, thereby achieving high-concurrency inference.

\begin{figure}
\centering
\begin{mdframed}[backgroundcolor=black!5,leftmargin=0cm,hidealllines=true]
\begin{lstlisting}
template <typename T>
class CardEstInferenceEngine {
// Load a CardEst model 
    bool loadModel(String modelPath);

// Validate model legitimacy
    bool validate();

// Initialize inference context
    void initContext();

// Featurize a SQL query into a vector
    FeatureVector featurizeSQLQuery(String sqlQuery);

// Featurize an abstract syntax tree into a vector
    FeatureVector featurizeAST(AbstractSyntaxTree ast);

// Perform CardEst inference using a feature vector
    double estimate(FeatureVector featVec);
};
\end{lstlisting}
\end{mdframed}
\vspace{-1em}
\caption{The APIs of Inference Engine}
\label{fig: infer_engine}
\end{figure}

\subsubsection{APIs for Integration with \BC's Query Processing}
The \IE offers two kinds of APIs for integration with \BC:
one for final estimation (via the \textsf{estimate} interface) and the other for featurization (via the \textsf{featurizeSQLQuery} and \textsf{featurizeAST} interfaces).
The final estimation is contingent upon the feature vector from the featurization phase. 
The specific inference algorithm of each \CE model can be implemented by their own probability calculations or matrix operations in this interface.

The featurization interface is orthogonal to the inference algorithms.
Its primary role is to capture the features of \BC's query-related data structures.
To facilitate this, two APIs are provided for \BC: 
One for featurization of SQL queries and the other for featurization of the abstract syntax tree (AST) produced by the \BC's analyzer.
SQL-based featurization is designed for easy integration with emerging inference algorithms developed by the research community, as these algorithms usually develop featurization methods directly based on SQL queries. This interface is handy for rapid proof-of-concept evaluation.
Alternative featurization, which leverages AST structures, is more effective in extracting richer features, including syntactic structures.
Notably, modern database systems often utilize non-standard in-memory AST structures, leading to a lack of portability.
Therefore, to harness the advantages of learned cardinality estimation using AST-based featurization in different systems, specialized inference algorithm implementations aligned with their specific AST structures must be developed. 
This customization would greatly help maximize the utility of the learning-based cardinality estimators, facilitating more accurate estimation for the system.

\subsection{ModelForge Service}
The \MS is designed to encapsulate the training algorithms of learning-based \CE models into a standalone service, which facilitates the automatic training process for different models.
The decision to develop a standalone service for model update on upstream data is driven by two factors: 1) Continuous sampling and training directly on the data stored in the storage layer would be resource-intensive and might risk impairing the performance of online customer queries. 2) This dedicated service enables \FS to easily incorporate the latest training algorithms of \CE models from the research community. 

Within \MS, there are two main tasks: routine training for \COUNT \CE models and occasional fine-tuning for \COUNTD \CE models. The regular training of \COUNT \CE models per table involves structural learning of Bayesian Networks using the Chow-Liu tree algorithm \cite{chow1968approximating}, followed by parameter learning based on the Expectation Maximization (EM)~\cite{do2008expectation} applied to the discovered structure. In addition, the fine-tuning of \COUNTD \CE models for individual columns is designed to adjust the pre-trained \LNE model to some specific columns where the original parameters are less effective. This process allows the model to learn the unique features of the specific columns and improve the estimation of their NDVs.

The initial \COUNT models for existing database tables in \BC are trained on the online sampled data, with the sampling process scheduled during low-activity periods of the \BC cluster.
Whenever the data updates come, \BC's Data Ingestor signals the service with related information on data consumption, which is essential for model updates. The information for Apache Hive includes table schema, data format, and location, while for Apache Kafka, it contains topic names, data formats, and offset details.
The retraining process begins after gathering sufficient data from upstream sources. The updated model is then saved in a specific location in the cloud storage, making it accessible later for \ML. The data used for training is automatically deleted after a set period.
To further improve the effectiveness of learned estimators, \MS supports training for individual table shards, particularly when there is significant variation in data distribution across different shards. This process involves obtaining the shard keys and functions for the training service, segmenting the training data accordingly, and then training local models for each shard.

\subsection{Auxiliary Modules} \label{sec: facilities}
This subsection introduces \FS' auxiliary modules, including \MPP and \MM.

\subsubsection{Model Preprocessor}
This module performs data preprocessing in \BC's query analyzer and optimizer, facilitating the training and inference processes of different \CE models. The key steps consist of column selection, type mapping, and join collection.
The first step, column selection, involves excluding columns with complex types such as \texttt{Array} and \texttt{Map}, which are beyond the processing capabilities of current models. The second step of type mapping is developed to convert the database type of each selected column into compatible types with machine learning algorithms. For example, machine learning typically uses types like \texttt{Binary}, \texttt{Categorical}, and \texttt{Continuous}.
The results after the above two steps are recorded in a system table named \textsf{model\_preprocessor\_info}. The \MS then accesses this table to retrieve essential information (such as which columns to read) and uses the type mapping to conduct the actual training.

The final step, join collection, involves gathering join patterns with \BC's analyzer. This process is critical because data warehouses' customers are not required to define the relationships of the primary key to the foreign key (PK-FK) during the table creation. 
For multi-table \CE models, the join pattern serves as an essential input for the training process, as these models are required to capture the joint distribution of the join keys.

\subsubsection{Model Monitor}
To ensure stability, \FS includes a \MM to oversee the quality of \CE models trained by \MS, ensuring that inferior models do not hinder the query processing of \BC.
Following the evaluation method from \cite{cardestbench}, \MM automatically generates queries for \COUNT and \COUNTD estimation with multiple predicates. These queries are then executed by \BC to obtain true cardinalities, enabling the \MM to make estimations and compute Q-Errors for \CE models. 
Models are retained only if their Q-Error is below a certain threshold. If the models cannot meet this threshold after several trials, \FS reverts to traditional methods for estimating the cardinality of the affected tables, ensuring consistent performance and reliability.

Note, \MM is configured to monitor only the single-table \COUNT and \COUNTD models, and the multi-table \COUNT models are excluded. This is because \BC can't afford the substantial computational resources that are needed to calculate true join sizes in Q-Error computation.
Given that the multi-table model (i.e., \FJ) used by \FS relies on single-table models for join-size estimation, monitoring the performance of single-table models (i.e., Bayesian Networks) indirectly contributes to the oversight of multi-table models. 
This approach ensures efficient resource utilization while maintaining the accuracy and reliability of the learned estimators in \FS.

%% file: sections/adapt.tex
\section{Model Integration}
\label{sec: model_adapt}
The training algorithms of all \CE models are deployed with \MS without disrupting \BC's query processing. This section focuses on the integration of model inference.
Given the constraints of \textsf{Python} in multi-threading, especially due to the Global Interpreter Lock (GIL)~\cite{beazley2010understanding}, \FS implemented inference algorithms with C++ to enhance system efficiency.

\subsection{The Single-Table \textsc{COUNT} Model}

 \begin{figure}[t]
    \centering
    \includegraphics[width=8cm]{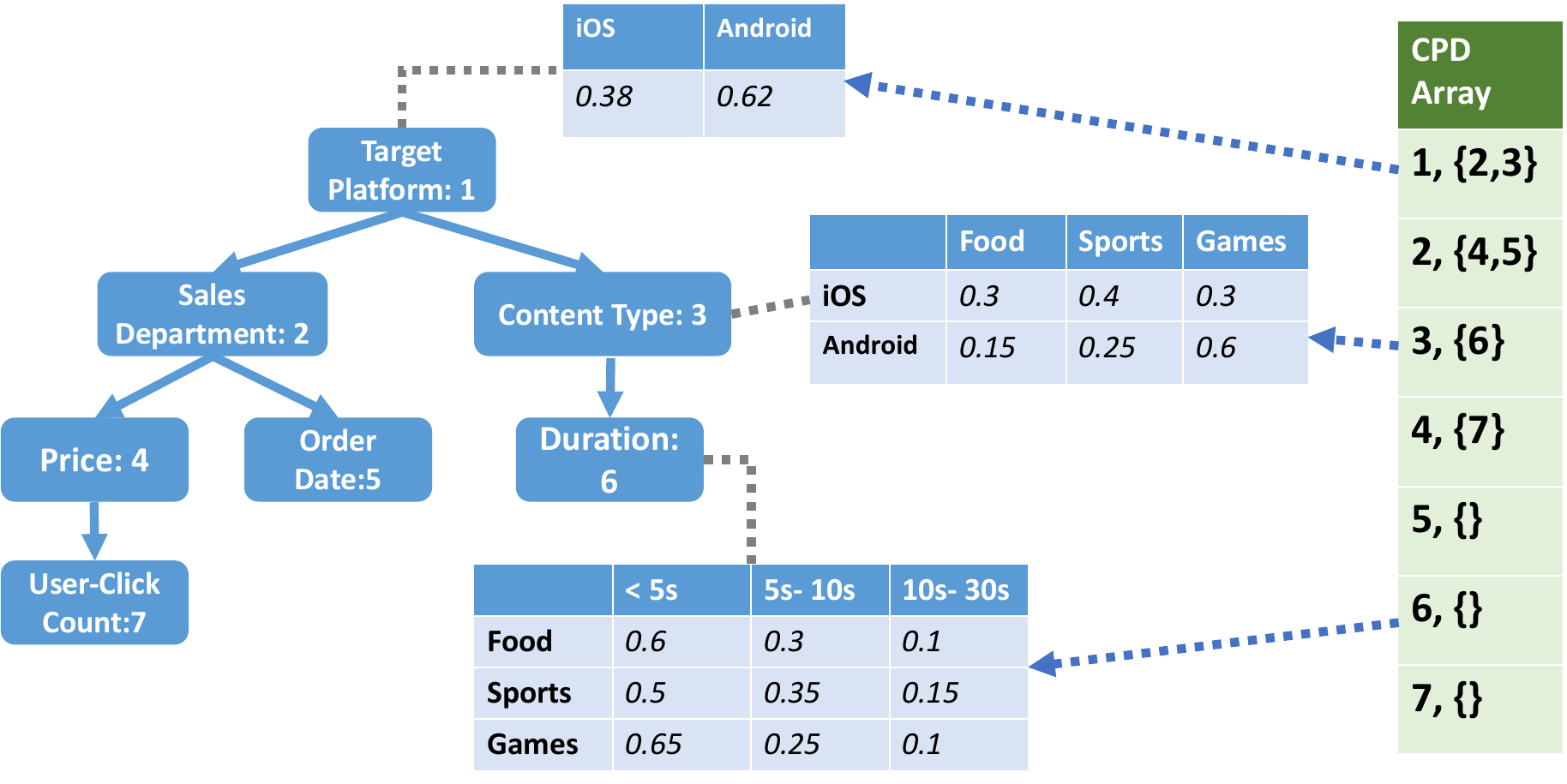}
    \vspace{-1em}
    \caption{The Single-Table BN Model with its \textsf{CPD}s}
    \label{fig: bn_ad_model}
\end{figure}

We employ the tree-based Bayesian Networks (BNs)~\cite{agrum2017} as our single-table model for \COUNT estimation. 
A tree-based BN is a probabilistic graphical model representing a set of variables and their conditional dependencies via a tree structure. \FS utilizes this model to capture the joint probability distributions across table columns.
In this model, each node represents a random variable (corresponding to a table column in the context of \CE). Each edge denotes the conditional dependencies between these variables (corresponding to the correlation across table columns), which are captured by structures known as the conditional probability distributions (\textsf{CPD}s). 
In \FS, the \textsf{CPD}s are represented as one-dimensional (1D) vectors or two-dimensional (2D) matrices.

Figure~\ref{fig: bn_ad_model} illustrates a distilled BN model trained from a business table in an online \BC cluster for analyzing the advertising placement strategy.
The columns of the table are structured as a tree, with \textsf{Target Platform} as the root node and \textsf{Content Type} as a dependent child node.
The adjacent \textsf{CPD}s depicts the dependence between those columns. 
The \textsf{CPD} of \textsf{Target Platform} is a 1D vector, while the \textsf{CPD} of \textsf{Content Type} and \textsf{Target Platform} are 2D matrices to capture the probabilistic relationships between the columns.

The inference algorithm for single-table cardinality estimation is based on the method of variable elimination (\textsf{VE})~\cite{koller2009probabilistic}.
Applying \textsf{VE} for BNs should avoid potential data races in multi-threaded environments. To address this issue, \FS proposes two key techniques to the \textsf{initContext} interface, which are:

\begin{enumerate}

 \item \textbf{Root Identification}: The \textsf{VE} algorithm initiates at the root node and moves towards the leaves, involving a probability message passing down the tree to update distributions at each level. 
In order to prevent data races between query threads and enable high-concurrency inference invocations in multi-threaded query processing environments, we can achieve this by identifying the root of each model and making it immutable in the \textsf{initContext} interface, instead of using a global lock for the entire model.
\item \textbf{\textsf{CPD} Indexing}: The execution of single-table model inference, conducted via the \textsf{estimate} interface, requires frequent accesses to the values of probabilities in the \textsf{CPD}s. 
The \textsf{CPD}s are often located in the tree structure's nodes. However, it is inefficient to perform repeated traversal of the tree to access \textsf{CPD}s during inference.
To address the issue, the tree structures with CPDs are transformed into an array indexed according to the topological order of nodes. This array also records the information of nodes' children for convenient reference.
As illustrated in Figure~\ref{fig: bn_ad_model}, the root node \textsf{Target Platform} is assigned an index of $1$,  with subsequent numbers reflecting the topological order of its children. Meanwhile, the leaf node \textsf{Duration} is assigned an index of $6$ without any children. This indexing mechanism, integrated within the \textsf{initContext} interface, enables direct access to any CPD by its index, thereby eliminating the need for repeated traversal through the tree structure.
\end{enumerate}

\subsection{The Multi-Table \textsc{COUNT} Model}
\FS selects the \FJ model for multi-table join estimation. 
\FJ employs an approach that integrates single-table models to analyze join key distributions, which subsequently partitions the joint domain of these keys into discrete buckets (i.e., join-buckets). It effectively utilizes a factor graph model to encapsulate these keys within a probabilistic graphical paradigm, which further facilitates the computation of an upper bound on the join sizes. 
To effectively integrate \FJ's inference algorithm, it is important to invoke the \textsf{initContext} interface of each related single-table model during the initialization phase.
Besides, \FS develops two key techniques to facilitate the inference of \FJ:

\begin{enumerate}
\item \textbf{Join-Bucket Construction}: \MPP plays a crucial role in creating join-buckets for \FJ, essential for join-size inference on the factor graph. 
This process relies on two facilities: One is the join schema collected by the step of join collection, as mentioned in Section~\ref{sec: facilities}. The other one is the equi-height histograms that are built within the optimizer of \BC. 
Leveraging these resources, \MPP can construct join-buckets based on the joint value domain of all join keys.

\item \textbf{Distribution-Dimension Reduction}: A standard fact table with several join keys is common in business scenarios.
Inference on the factor graph needs to maintain the joint distribution of these join-keys. However, excessive join-keys in a single table can lead to high distribution dimensionality and increase inference complexity.
The solution of \FJ is to explore the causality patterns between these join-keys with a tree probabilistic structure, such that the dimensionality of the joint distribution can be effectively reduced.
To apply this method, \FS leverages the same training procedure of the Chow-Liu algorithm in \MS. 
This approach greatly reduces the inference complexity of join-size estimation of fact tables by reducing the distribution dimension.
\end{enumerate}

\subsection{The \textsc{COUNT-DISTINCT} \CE Model}
\FS has chosen \LNE as its NDV estimator, which employs a seven-network layer training on the general feature called ``frequency profile'' of NDV estimation. The ``frequency profile'' is a compact representation of the frequency distribution of distinct values calculated based on a sample of column data.
Although training \LNE from the ground up is time-consuming, it usually does not require retraining when facing new workloads. Upon completion of the training, the model weights can be stored in the cloud storage.

To integrate \LNE into our framework, \FS loads the neural network architecture of \LNE into memory during the startup of \BC. Meanwhile, the \MM is responsible for loading the \LNE model parameters through the \textsf{initContext} interface.
Unlike the frequent loading of Bayesian Networks' parameters, the \LNE model weights are loaded with a relatively lower frequency because of the models' workload-independent properties.

The featurization process of \LNE involves computing an important feature called the "frequency profile". This task is computationally intensive and can significantly impact the overall inference performance of \LNE. In optimization scenarios where real-time NDV estimation is necessary, it is critical to create the "frequency profile" to provide accurate NDV estimates to \BC efficiently. Section~\ref{sec: lne_application} provides insights into a specific estimation scenario where we apply \LNE's estimation and refine the computation of the "frequency profile". Once the feature extraction is complete, the main inference computation within the estimate interface involves matrix multiplication operations on the neural network.

%% file: sections/applications.tex
\section{Enhanced Query Optimization} \label{sec: applications}
This section shows how \BC leverages the accurate cardinality estimates from \FS to enhance its query optimization. While the discussion is limited to only two optimization scenarios, \FS has the potential to offer broader benefits.

\subsection{The Cases for Materialization Strategy}
We enhance \BC's previous strategy by introducing a multi-stage reader approach. In contrast to the single-stage reader that retrieves columns and applies predicate filtering in one pass, this approach incrementally constructs tuples through sequential filtering and appending columns. To support different workloads, \BC utilizes different strategies. This approach, together with the integration with \FS, enables the materialization strategies to reduce I/O costs during query processing significantly. 
At last, we will discuss how join-order selection affects the efficiency of materialization strategies.

\subsubsection{Column-Order Selection in Multi-Stage Reader}
In the multi-stage reader, the order to access the required columns is crucial. Specifically, it is beneficial to prioritize highly selective columns to minimize I/O overhead in subsequent stages.
The strength of learned \CE models is their capability to capture cross-column correlations, which is challenging for traditional histogram-based approaches.
Although the independent storage of each column in \BC might suggest that cross-column correlations are negligible, this assumption is incorrect in certain cases.

\begin{example*}
\textit{Assume, for contradiction, that the selection order of columns, despite their cross-column correlations, has no substantial effect on I/O overhead. Consider an instance where filters are set as \textsf{col1 > 0 AND col2 > 0 AND col3 > 0}, with \textsf{col1} independent of \textsf{col2} and \textsf{col3}, which are strongly correlated (\textsf{col2 = col3 + 2}). This assumption leads to \textsf{prob(col2 > 0) <= prob(col2 > -2) = prob(col3 > 0)}. Assuming \textsf{prob(col1 > 0)=0.7}, \textsf{prob(col2 > 0)=0.6} and \textsf{prob(col3 > 0) = 0.8}, a straightforward selectivity estimation would prioritize \textsf{col2->col1->col3}. However, \textsf{prob(col2 > 0 AND col3 > 0) = prob(col2>0) < prob(col1 > 0)}, suggesting that reading \textsf{col2} and \textsf{col3} before \textsf{col1} would minimize I/Os.
This contradicts our initial assumption, proving that the order of column access, especially considering cross-column correlations, critically impacts I/O overhead. Therefore, accounting for these correlations and optimizing the access order of columns is essential to maximize the effectiveness of learned \CE models.}

\end{example*}

To exploit the power of learned \CE, it is necessary to estimate the selectivity of column combinations along with their corresponding predicates. This incurs additional overhead due to the need to enumerate all possible orders of read columns. 
To mitigate this overhead, we impose the constraints on the enumeration process. Specifically, we early-stop the enumeration if the selectivity of the current generated combination exceeds a predefined threshold. This practice allows us to simplify the enumeration process while leveraging the benefits of learned \CE models.

\subsubsection{Dynamical Decision of Reader Selection}
No single materialization strategy is universally optimal due to the diversity of analytical workloads. While the multi-stage reader effectively reduces unnecessary readingi? I/Os and performs well in most cases, it has been observed that for some queries, the multi-stage reader might incur more I/Os than its single-stage alternative.
These cases often occur when the query predicates are non-selective, requiring the multi-stage reader to go through a significant part of the dataset. 
This often leads to increased maintenance overhead, as the multi-stage needs to maintain essential processing information across the stages.
To select the best materialization strategy for a query, \BC utilizes \FS's cardinality estimates to calculate the query's overall selectivity. 
If the overall selectivity of the query is high, \BC will choose the single-stage reader approach. Alternatively, if the overall selectivity is low, \BC will default to the multi-stage reader approach.
The single-table \CE model (i.e., tree-based BN) that we employ proves rather efficient for queries with multiple \textsf{AND}-ed predicates, due to its inherent modeling of joint probability distributions across columns. In practice, \FS uses the \textit{inclusion-exclusion} principle to transform \textsf{OR}-ed queries to \textsf{AND}-ed formats before calculating selectivities.

\subsubsection{Join-Size Estimation}
The effectiveness of materialization strategies for multi-table join queries is heavily influenced by two factors: the size of the join and the order in which the joins are performed. 
The size of the join affects the performance-critical decision of whether to materialize tuples before or after applying the join operation in the query plan~\cite{abadi2006materialization}.
On the other hand, the join order significantly impacts the materialization overhead that occurs during join processing, especially when large tables are involved in early join operations. Therefore, it is crucial to estimate the join size and infer the join order to minimize materialization overhead and improve efficiency.
In Section~\ref{sec: choice}, we discussed how \FS effectively estimates join size by utilizing \FJ,  which ensures that the assumption of join-uniformity is avoided, allowing for more accurate estimates.
With \FJ, \FS has already shown considerable promise in reducing materialization overhead and enhancing \BC's join processing efficiency.
Notably, the enhanced accuracy in join size estimation enables \BC to optimize join order, significantly reducing the amount of intermediate results that need to be materialized.

\subsection{The Cases for Aggregation Processing}
\label{sec: lne_application}
This subsection discusses how \FS utilizes \LNE estimator to optimize the aggregation processing in \BC and how to deal with the cases where the estimator may underperform.

\subsubsection{How \LNE can help?}
During query processing, \textit{resizing} the aggregation hash table can have an impact on query performance, especially regarding memory management. Therefore, it is essential to have an accurate estimate of the initial size for the hash table. Underestimating the initial size can lead to frequent resizing due to rapid saturation, which can degrade performance. Overestimating the initial size can lead to unnecessary disk spillover or suboptimal memory utilization.

The estimation of hash table size can benefit from \FS's learned NDV estimator (i.e., \LNE), which adapts to the diverse data distributions in different scenarios.
This approach differs fundamentally from traditional approaches that use statistics collection in optimizers to calculate NDV estimations for different columns beforehand.
A notable challenge in hash-table size prediction for aggregation processing is that the data amount of aggregated columns can be heavily influenced by ad-hoc predicates, making the pre-computation of NDVs impractical.
To estimate the hash table size using \LNE, \FS needs to construct a key feature (i.e., ``frequency profile'') in the featurization interface as mentioned in Section~\ref{sec: model_adapt}.
To build this feature, \ML loads a small sample (nearly 10 million rows) for each table and converts them into a \textsf{DataFrame} format with a high-performance \textsf{C++} library. The \textsf{DataFrame} is a mutable two-dimensional table supporting different data types and labeled axes for in-memory computation.
This structure enables efficient filtering and calculation of the ``frequency profile''.
Once the ``frequency profile'' has been calculated, \FS facilitates \LNE's actual inference through the \textsf{estimate} interface.

 \subsubsection{Calibration.}
 \label{sec: algo_ob}

While the \LNE estimator offers \BC high-accuracy estimation in most cases, it may underestimate NDV when the true number of distinct values in a column is exceptionally high.
To address this issue, \FS has developed a calibration protocol to fine-tune the estimation of the \LNE model. 
If the \MM detects poor NDV estimates with large Q-Errors of some columns, it will initiate a fine-tuning procedure for the columns that have been identified as problematic in the \MS.
Technically, this procedure augments the original \LNE's training dataset with sampled data from the problematic columns, alongside additional synthetic data characterized by high NDVs. The model is then retrained from the last checkpoint, which is applied in general use cases.
For scenarios with exceptionally high NDVs, the retraining process uses a relatively small learning rate and imposes more penalties for underestimation cases.
Once the fine-tuning is completed, the refined neural network parameters are saved in the cloud.
Later, \ML makes use of these parameters to build a calibrated model. It is worth mentioning that these updated parameters are specifically trained to adjust and calibrate only the problematic columns.

%% file: sections/experiments.tex
\section{Performance Evaluation}
\label{section: evaluation}
This section evaluates \FS's effectiveness with workloads from academic and industrial datasets.
We begin our evaluation by analyzing how \FS enhances the query processing of \BC across different workloads. This end-to-end evaluation proves the practical benefits of the proposed framework. Then, we analyze this improvement from the system's perspective, focusing on metrics like reading I/Os. Finally, we examine the accuracy of \FS's cardinality estimates from the algorithm's perspective and provide further observations on \FS's models.

\subsection{Experimental Setup}
All experiments are conducted on a large-scale \BC cluster with the specifications detailed in Table~\ref{tab: setup}.
 
\noindent\underline{\textbf{Datasets.}} We utilize three datasets:
IMDB~\cite{leis2015good}, STATS~\cite{cardestbench} from the academic community, and AEOLUS from our internal business scenario. 
The original sizes of the two academic datasets are relatively small, so we scale them to $1$TB using the method proposed in~\cite{hizeroshot2022}. This scaling method preserves the original data distribution, making it easy to calculate the actual cardinality. 
 
\noindent\underline{\textbf{Workloads.}} 
 We choose the JOB-LIGHT~\cite{MSCN} 
and STATS-CEB~\cite{cardestbench} workloads for the IMDB and STATS dataset as they are the latest benchmark sufficiently complex to evaluate \CE methods.
To evaluate the performance of aggregation processing, we manually extend the original workloads by adding queries that reflect practical analytical usage. These queries, together with the original ones, created new workloads called JOB-Hybrid and STATS-Hybrid. For instance, the aggregation queries for the STATS dataset include an average score of user posts and several comments per post by year.
For the AEOLUS dataset, we use a workload called \textsf{AEOLUS-Online} from an online business scenario. The workload includes five business tables and features a mix of various join and aggregation queries. The statistical information of the three workloads is presented in Table~\ref{tab: workload-stat}.

\begin{table}[t]
	\caption{Machine and Cluster setup.}
	\scalebox{0.9}{
		\begin{tabular}{c|c}
			\hline
			\hline
			CPU & Intel(R) Xeon(R) Gold 6230 \\ 
			& (CPU @ 2.10GHz and 75 cores) \\\hline
			Memory & 300~G  \\\hline
			Network & 10Gbps Ethernet \\ \hline
			OS & Debian 9 (Linux Kernel Version 5.4.56)  \\ \hline
			Cache & 55M shared L3 cache  \\ \hline\hline
			
			Server & 1  \\ \hline
			Compute-Worker & 8 \\ \hline
			Ingestor-Worker & 8 \\ \hline
			
		\end{tabular}
	}
	\label{tab: setup}
\end{table}

\begin{table}[t]
	\caption{Workload Statistics.}
 \vspace{-1em}
	\scalebox{0.73}{
		\begin{tabular}{c|ccc}
			\hline
			\rowcolor{mygrey}
			\textbf{} & \textbf{\textsf{JOB-Hybrid}} & \textbf{\textsf{STATS-Hybrid}} & \textbf{\textsf{AEOLUS-Online}} \\ \hline
			\# of queries & 100 & 200 & 200\\ \hline
			\# of join templates & 23 & 70 & -\\ \hline
			\# of joined tables & 2-5 & 2-8 & 2-5 \\ \hline
			\# of \textsf{group-by} keys & 1-2 & 1-2 & 2-4 \\ \hline
			range of true cardinality & $9\cdot10^3$ --- $9\cdot10^{12}$ & $5.2\cdot10^4$ --- $4.4\cdot10^{12}$ & $7\cdot10^3$ --- $4.7\cdot10^{11}$ \\ \hline  
   \parbox{3cm}{\# of queries hit the max joined-table} & 31 & 6 & 7\\ \hline
   \parbox{3cm}{\# of queries hit the max group-by key} & 11 & 13 & 50\\
   \hline
		\end{tabular}
	}
	\label{tab: workload-stat}
\end{table}

\subsection{Query Latency}

We compare end-to-end query performance between two traditional \CE methods and \FS on the three workloads. The first method leverages sketch-based algorithms (Histogram and HyperLogLog) with pre-computed sketches for each dataset. The second is sample-based, akin to AnalyticDB's approach~\cite{zhan2019analyticdb}. We standardize sample rates and the degree of parallelisms across methods for a fair comparison and disable query caching in all experiments to ensure unbiased results.
The results are plotted in Figure~\ref{fig: query_perf}, with latency normalized against the highest value in each plot.
The figure illustrates the 50th, 75th, 90th, and 99th percentile query latency across three workloads.
For each workload, \FS demonstrates the optimal latency almost at all quantiles. This efficiency is due to its accurate cardinality estimates, which are applied in performance-critical execution paths in \BC, particularly in materialization strategies and join-order selection.

At the lower latency quantiles, the sketch-based method performs better than the sample-based method, while \FS exhibits comparable efficiency. 
The sample-based approach shows suboptimal performance due to the need for predicate computation during real-time sampling. This process incurs significant overhead in the cardinality estimation stage, a limitation not present in the other two methods.
At higher quantiles of latency, \FS outperforms traditional methods, especially in terms of the P99 latency of the \textsf{STATS-Hybrid} workload, improving it by at least 30\%. This improvement is attributed to \FS's selection of lightweight models, which provide high-accuracy cardinality estimation and benefit from efficient inference procedures. 
The STATS workload presents a complex data distribution, which poses a challenge for traditional methods to deliver accurate cardinality estimates. Therefore, the marked enhancement of the STATS workload is due to \FS's ability to overcome this challenge.

\begin{figure*}[ht]
  \centering
  \subfigure[\vspace{-5pt}\textsf{JOB-Hybrid}]{
    \label{fig:subfig:imdb_latency}
    \includegraphics[scale=0.3]{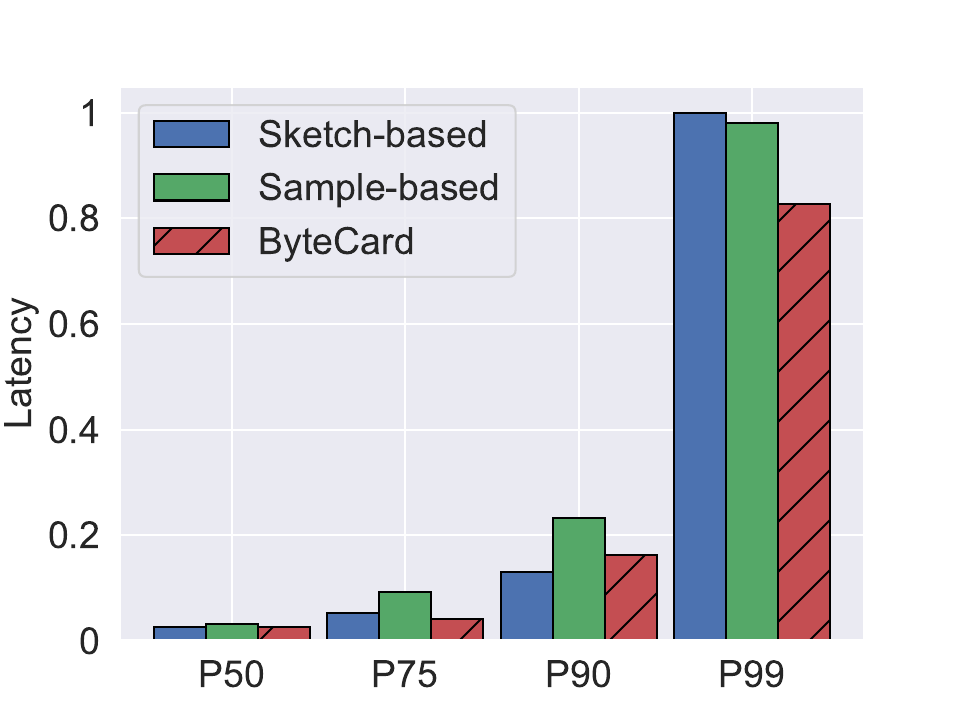}
    }
  \subfigure[\vspace{-5pt}\textsf{STATS-Hybrid}]{
    \label{fig:subfig:stats_latency}
    \includegraphics[scale=0.3]{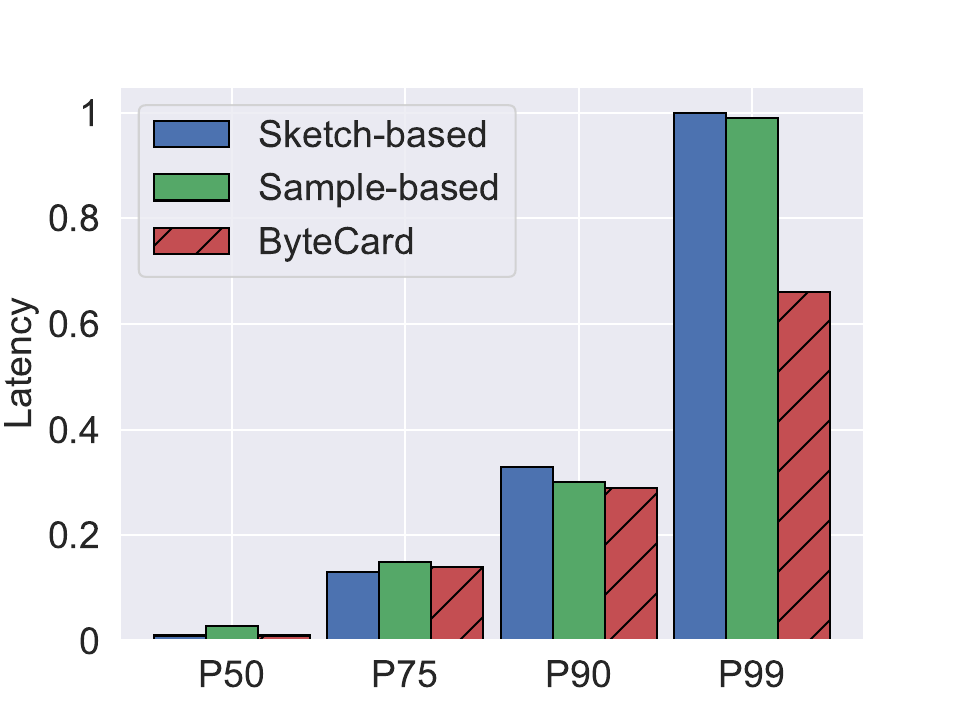}
    }
   \subfigure[\vspace{-5pt}\textsf{AEOLUS-Online}]{
    \label{fig:subfig:aeolus_latency}
    \includegraphics[scale=0.3]{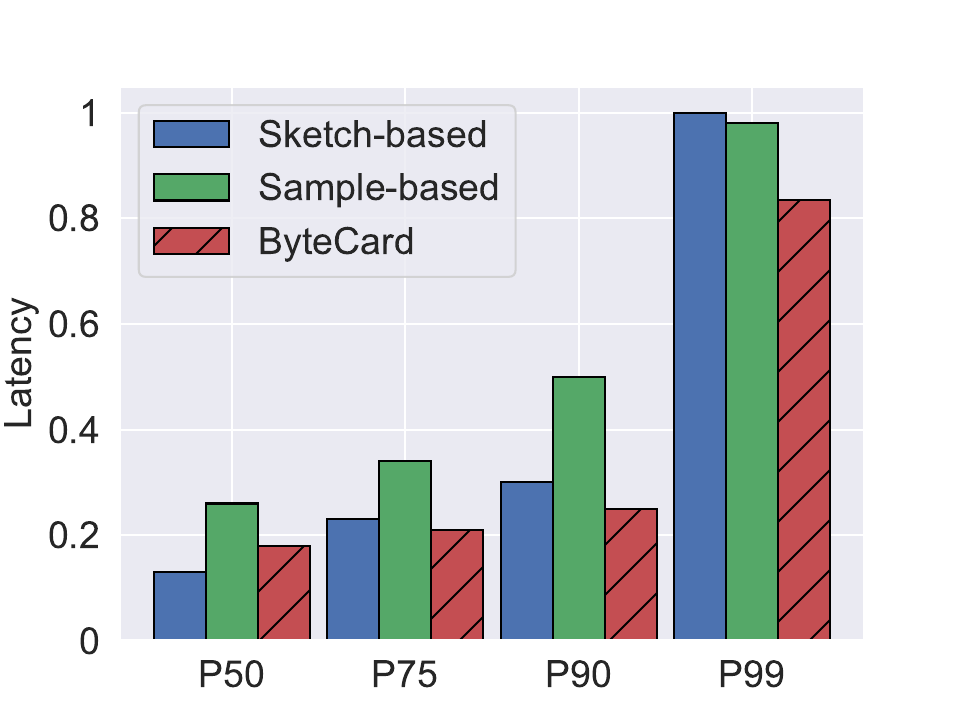}
    }
  \vspace{-1em}
  \caption{Query Latency Across Different Workloads}
  \vspace{-1em}
  \label{fig: query_perf}
\end{figure*}

\subsection{System Analysis}

\renewcommand{\subfigcapskip}{-0.3em}
\renewcommand{\subfigbottomskip}{-0.3em}
\begin{figure}[ht]
  \centering
  \vspace{-0.35em}
  \subfigure[Reading I/Os]{
    \label{fig: reduced_io}
    \includegraphics[scale=0.27, trim=5 0 10 30,clip]{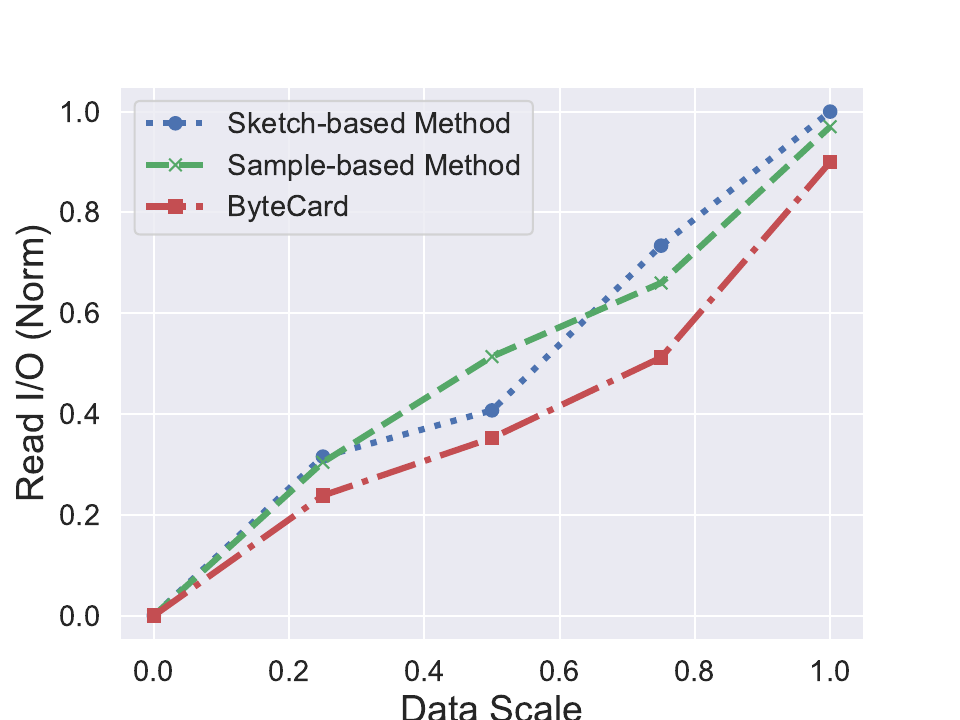}
    }
    \hspace{-1.6em}
  \subfigure[Resizing Frequency]{
    \label{fig: reduced_resize}
    \includegraphics[scale=0.27, trim=5 0 10 30,clip]{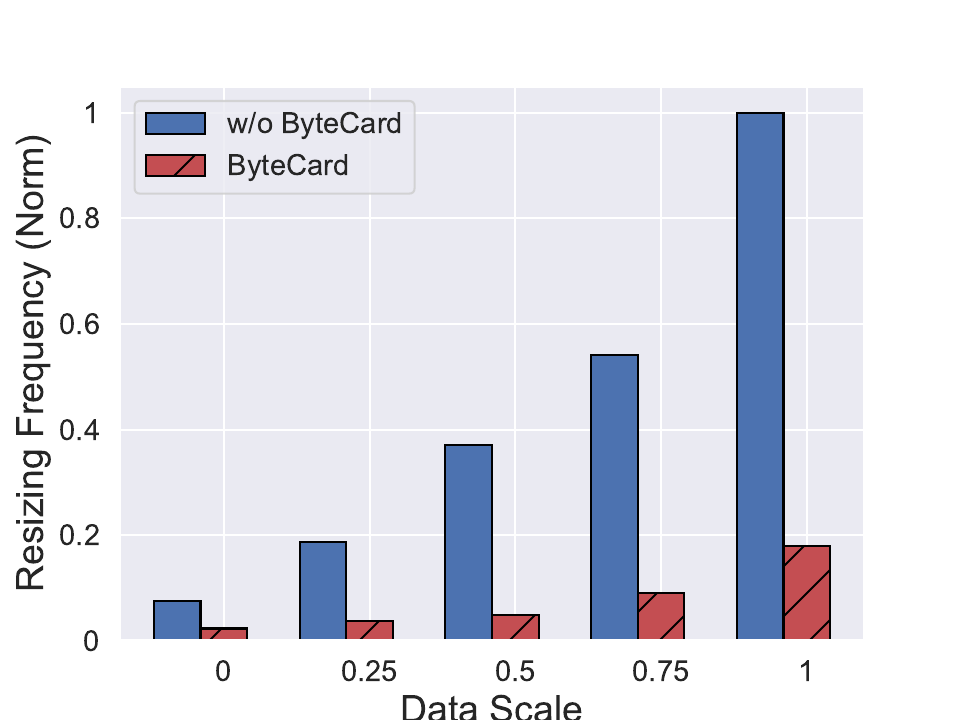}
    }
    \vspace{-1.5em}
  \caption{\small{The Observed System Metric Across Different Data Scales}}
  \label{fig: query_perf}
\end{figure}

\renewcommand{\subfigcapskip}{-1em}
\renewcommand{\subfigbottomskip}{-1em}
We examine the system's perspective to see how \FS improves query latency across different workloads.
In our experiment, we split the STATS and AEOLUS datasets, train models at each scale, and evaluate \FS's impact on reducing reading I/Os and hash table resizing during aggregation processing.

\begin{figure*}[htb]
\vspace{-0.2em}
  \centering
  \subfigure[\textsf{JOB-Hybrid}]{
    \label{fig:subfig:imdb_qerror}
    \includegraphics[scale=0.3]{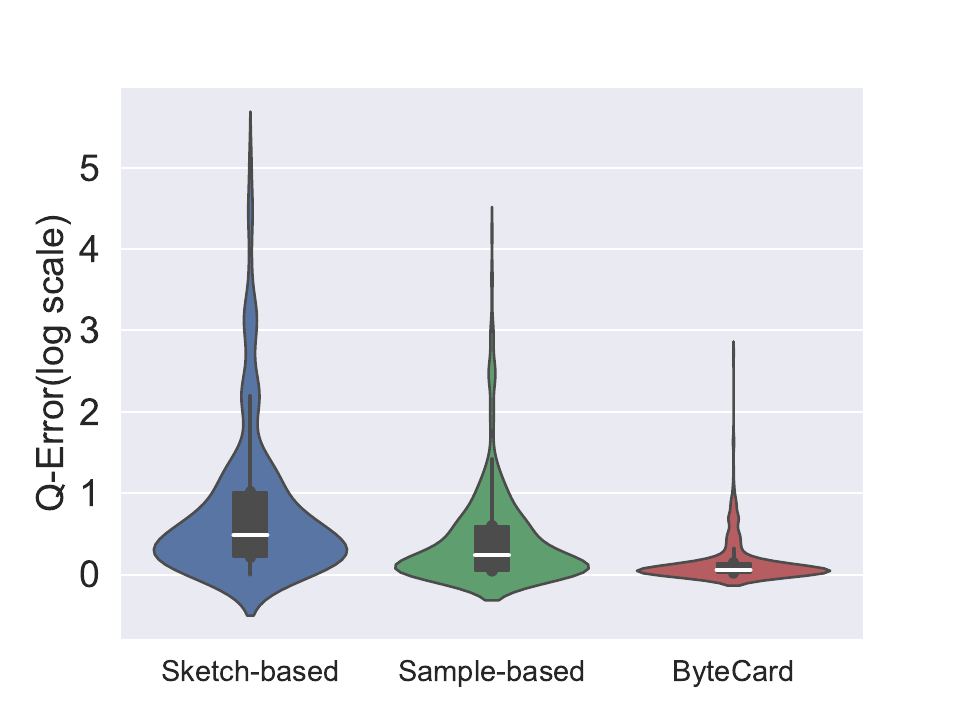}
    }
  \subfigure[\textsf{STATS-Hybrid}]{
    \label{fig:subfig:stats_qerror}
    \includegraphics[scale=0.3]{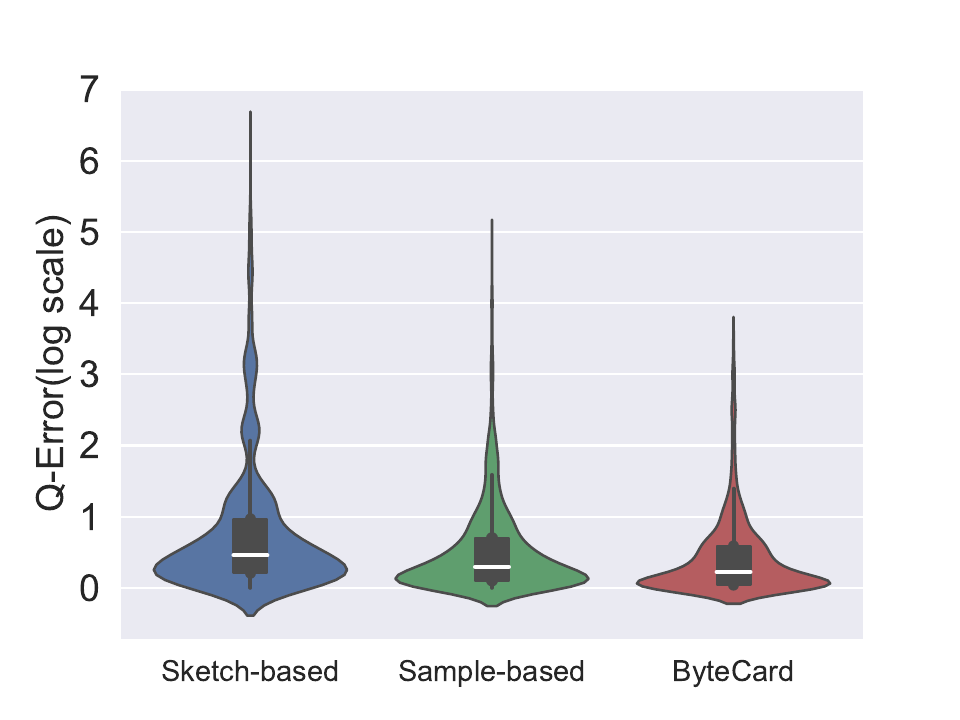}
    }
   \subfigure[\textsf{AEOLUS-Online}]{
    \label{fig:subfig:aeolus_qerror}
    \includegraphics[scale=0.3]{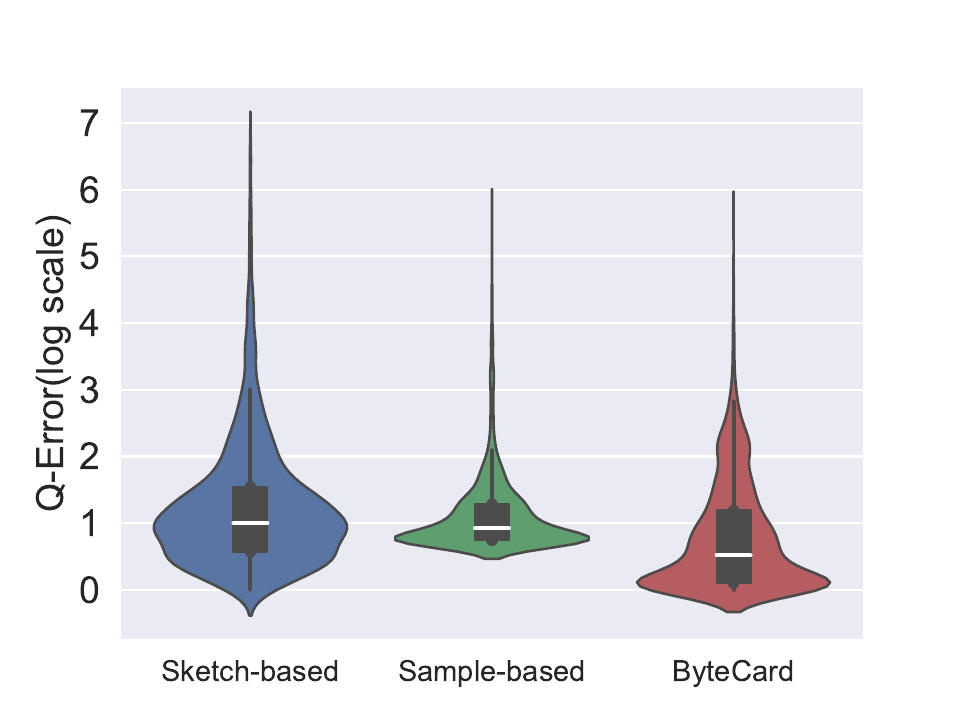}
    }
  \caption{The Observed Algorithm Metric Across Different Workloads}

  \label{fig: query_qerror}
\end{figure*}

\noindent\underline{\textbf{Reading I/Os.}} Figure~\ref{fig: reduced_io} illustrates the reading I/Os for processing the \textsf{STATS-Hybrid} workload across STATS' different scales, with the results normalized to the observed largest size. In smaller data scales, the sketch-based method 
 aids \BC's materialization strategy in reducing reading I/Os more effectively than the sample-based method, owing to its relatively accurate estimates. However, as the data scale enlarges, the sketch-based method's performance deteriorates due to its reliance on simplified assumptions. In contrast, the sample-based method delivers more accurate estimations in larger data scales, benefiting from its flexibility and adaptability to changing data patterns.
Despite these advantages, both traditional methods are surpassed by \FS. \FS's superiority is attributed to its ability to capture cross-column and cross-table correlations, utilizing Bayesian Networks and \FJ.
Thus, \FS guides the materialization strategy for \BC to minimize reading I/Os more effectively.

\noindent\underline{\textbf{Resizing Frequency.}} Figure~\ref{fig: reduced_resize} displays the frequency of hash table resizing during the aggregation processing of the AEOLUS dataset at different scales. 
The sketch-based method, specifically HyperLogLog, fails to provide NDV estimation effectively in such a dynamic scenario; 
the sample-based method has significant overhead due to its real-time requirement to evaluate query predicates. 
Considering the given limitations, neither of the methods is considered appropriate for scenarios involving aggregation processing.
As a result, our analysis is limited to evaluating whether enabling \FS in \BC would reduce resizing frequency.
The results revealed by Figure~\ref{fig: reduced_resize} emphasize the efficiency of \LNE's integration in significantly reducing the necessity for hash table resizing during the aggregation processing of \BC, showcasing its superior performance and adaptability.

Note, \LNE's workload-independent nature eliminates the requirement of separate model training for each dataset scale. As the data scale enlarges, resizing frequency rapidly increases in the absence of \FS. In contrast, by utilizing \LNE's estimates, \BC shows remarkable effectiveness in significantly reducing the frequency of hash table resizing, even amidst escalating data scales. This highlights the effectiveness of \LNE in dynamically adjusting to various data volumes, thereby enhancing \FS's effectiveness in memory management during aggregation processing.

\subsection{Algorithmic Observations}
In this set of experiments, we delve into an algorithmic perspective to evaluate the estimation accuracy of \FS. Then, we examine the resource consumption of the models employed by \FS, focusing on model size and training time.

\noindent\underline{\textbf{Q-Error.}} Figure~\ref{fig: query_qerror} presents the Q-Errors using violin plots.
For all the distributions of Q-Errors, we can see that most are concentrated on smaller values indicated by the width of the violin.
The white line in the middle of the box inside each violin represents the median of Q-Error, while the black rectangle shows the interquartile range (the middle 50\% of the values).
When it comes to the critical aspects of the plots, such as the median of Q-Errors and interquartile range, traditional methods and \FS exhibit different characteristics.

The median Q-Error of \FS across all the workloads achieves the lowest value among the three evaluated methods. Its interquartile range is also relatively lower than the two traditional methods.
For \textsf{JOB-Hybrid}, the sketch-based method's median Q-Error is comparable to that of the sample-based method but exhibits poorer performance at higher quantiles. For \textsf{STATS-Hybrid}, the sketch-based method's performance deteriorates further, attributable to STATS's more complex data distribution and the larger query space of STATS-Hybrid. In the case of AEOLUS-Online, the sample-based method demonstrates limited robustness, primarily due to the complexity of business queries, which hinders accurate estimates from small samples.
Interestingly, in some cases, the sample-based method demonstrates better Q-Errors than the sketch-based method. Yet, it does not translate into superior end-to-end query performance. This paradox arises because the improved cardinality estimation of the sample-based method comes at the expense of increased estimation overhead. This result emphasizes why \FS prioritizes models that offer high-accuracy estimation and efficient inference. Owing to this strategy of model selection and the \IE abstraction for efficiently integrating the inference algorithms, \FS consistently achieves the best Q-Errors across all workloads.

\begin{table}[t]
\caption{Details of \FS's Models Per Table}
\vspace{-1em}
\begin{spacing}{0.9}
\scalebox{0.9}{
\begin{tabular}{llll}
\hline
\rowcolor{mygrey}
\textbf{Dataset} & \textbf{Method} & \textbf{Model Size} & \textbf{Training Time} \\ \hline
\multirow{3}{*}{IMDB} & BN & $3.6$ Mb & $2.5$ min \\ \cline{2-4} 
 & \FJ & $2.9$ Mb & $1.9$ min \\ \cline{2-4} 
 & \LNE & $256$ Kb & - \\ \hline
\multirow{3}{*}{STATS} & BN & $4.2$ Mb & $3.1$ min \\ \cline{2-4} 
 & \FJ & $2.3$ Mb & $0.8$ min \\ \cline{2-4} 
 & \LNE & $256$ Kb & - \\ \hline
\multirow{3}{*}{AEOLUS} & BN & $4.3$ Mb & $2.2$ min \\ \cline{2-4} 
 & \FJ & $3.6$ Mb & $1.7$ min \\ \cline{2-4} 
 & \LNE & $534$Kb & $57$ min \\ \hline
\end{tabular}
}
\end{spacing}
\label{tab: model_info}
\end{table}

\noindent\underline{\textbf{Model Details.}}
Table~\ref{tab: model_info} contains information on the average size and training time of \FS's models, including Bayesian Networks extracted from the \MS. 
The table also shows details about \FJ, including the size of the \textit{join-bucket}s and their construction time as the training time.
We can see both models maintain a compact size, below $5$ Mb. 
When integrated into \BC, these models result in a moderate increase in memory footprint, but this expansion does not impose a substantial resource burden on the system.
The \LNE model doesn't require recording of training time for IMDB and STATS datasets as it's workload-independent and needs only a single offline training session. Thus, the model size remains consistent across these datasets.
While \LNE is generally effective, it faces challenges with columns in AEOLUS's tables with exceptionally high NDVs. To mitigate this, a calibration protocol involving model fine-tuning is developed.
During the fine-tuning process in \MS, the learning rate is reduced which leads to slower convergence. Fine-tuning for a single problematic column can take up to an hour. Note, this does not compromise \BC's stability. This is because the fine-tuning process is executed in \MS, which does not impact \BC's query processing. Besides, if \MM identifies poor NDV estimates from \LNE models for specific columns, it instructs \BC to switch to a traditional NDV estimator. \FS only integrates a new \LNE model for estimating these problematic columns once \MM has validated the calibrated parameters.

%% file: sections/lessons.tex
\section{Lessons and Future Directions}

\noindent \underline{\textbf{Limitations of Learned \CE Methods:}}
While learned cardinality estimators have considerable promise in improving \BC, their deployment can sometimes lead to suboptimal performance owing to several factors.
Firstly, the complexity of data distributions and the diversity of query workloads across different scenarios pose substantial challenges to achieving accurate cardinality estimates. This means that no single model can perform well in all scenarios.
For example, BayesCard, which is one of the models adopted by \FJ, is prone to underestimate large true cardinalities in comparison to its alternatives~\cite{cardestbench}.
Therefore, we plan to further explore emerging approaches such as meta-learning~\cite{wu2021unified,hilprecht2021one,hizeroshot2022} to tackle the challenge.
Secondly, current estimators are limited by their focus on base tables, requiring research that synergizes ML with the Cascades framework~\cite{graefe1995cascades}, which is foundational in modern query optimizers.
The Cascades framework typically employs the \textsf{memogroup} \cite{lee2023analyzing}, a base abstraction for organizing logically equivalent query plans or expressions. 
We recommend constructing \CE models around the \textsf{memogroup}, which plays an essential role in exploring potential query plans.
Thirdly, achieving accurate cardinality estimation does not ensure optimal system performance. Other critical factors, like cost estimation, also influence the overall system efficiency. The intricacy of systems and the fluctuating nature of cloud environments~\cite{siddiqui2020cost,schad2010runtime} render accurate cost modeling of query plans difficult despite having accurate cardinality estimates. Our future work will explore more opportunities for ML to enhance \BC's query processing~\cite{suncostmodels2019,siddiqui2020cost}.

\noindent \underline{\textbf{Model Preference for ML-enhanced Components:}}
Unlike approaches like knob tuning~\cite{zhang2019end,van2017automatic,shen2023rover,ZhangWCJT0Z021}, materialized view recommendations~\cite{han2021autonomous,liang2019opportunistic}, and others that apply ML for external logical tuning~\cite{van2017automatic,shen2023rover,zhang2019end}, \FS is an ML-enhanced component embedded in the core of system, aiming at the physical (in-kernel) optimization. 
When it comes to physical optimization, it is important to establish different criteria for model selection than those used in learning-based logical tuning approaches.
Unlike logical tuning where ML serves mainly as an intelligent advisor, physical optimization demands direct integration into the system core, necessitating different considerations for model selection.
For model selection in physical optimization tasks, we suggest prioritizing models that can perform fast inference. 
This means preferring models that are both compact and accurate over more complex options, such as large language models (LLMs)~\cite{wei2022emergent,gpt3_2020}. 
Large models require extensive training and resources, which make them impractical for physical optimization tasks where efficiency and speed are crucial.

\noindent \underline{\textbf{Future Integration of More ML-Enhanced Components:}} 
Integrating ML-enhanced components into database systems requires a thoughtful design, as different models have their unique characteristics. In a tightly coupled architecture, both model training and inference processes should work in the kernel. However, this approach limits the framework's ability to evaluate emerging learned approaches. 
Besides, developing both training and inference algorithms using the same native language of system development requires significant engineering effort.
In this work, we showcase an engineering example of deploying learned cardinality estimators with \FS by integrating the training algorithms in a standalone service and embedding the inference algorithms in the system kernel.
Our ongoing work aims to deploy more ML-enhanced components, such as learned cost estimators, into our system. Unlike cardinality estimators, cost estimators usually employ query-driven approaches to improve their estimate performance. These models, such as XGBoost~\cite{saxena2023auto} and Elastic Net~\cite{siddiqui2020cost}, require runtime traces or query plan statistics for training. 
To integrate the training algorithms for the learned cost estimators, \MS can trigger the training process after retrieving query logs collected from \BC.
To integrate inference algorithms into our system, we need to avoid \textsf{Python} implementations that may negatively affect query performance. 
Fortunately, the interface provided by \IE standardizes the integration process of inference algorithms.
When loading cost models, we can follow the existing process for \CE models. However, the initialization and validation stages require customized developments for each model. 
It's important to identify immutable data structures to prevent any potential data race during query processing.

%% file: sections/conclusions.tex
\section{Conclusions}

\FS is a novel framework that aims to integrate learned cardinality estimators into ByteDance's data warehouse system, \BC.
The framework's design comprises \IE and \MS, which enable efficient model training and inference, thus improving query processing without overburdening \BC's computational resources.
This work paves the way for \BC's future integration of more ML-enhanced components.

%% file: bytecard.bbl

\begin{thebibliography}{63}


\ifx \showCODEN    \undefined \def \showCODEN     #1{\unskip}     \fi
\ifx \showDOI      \undefined \def \showDOI       #1{#1}\fi
\ifx \showISBNx    \undefined \def \showISBNx     #1{\unskip}     \fi
\ifx \showISBNxiii \undefined \def \showISBNxiii  #1{\unskip}     \fi
\ifx \showISSN     \undefined \def \showISSN      #1{\unskip}     \fi
\ifx \showLCCN     \undefined \def \showLCCN      #1{\unskip}     \fi
\ifx \shownote     \undefined \def \shownote      #1{#1}          \fi
\ifx \showarticletitle \undefined \def \showarticletitle #1{#1}   \fi
\ifx \showURL      \undefined \def \showURL       {\relax}        \fi
\providecommand\bibfield[2]{#2}
\providecommand\bibinfo[2]{#2}
\providecommand\natexlab[1]{#1}
\providecommand\showeprint[2][]{arXiv:#2}

\bibitem[Abadi et~al\mbox{.}(2009)]%
        {stonebraker2018c}
\bibfield{author}{\bibinfo{person}{Daniel~J Abadi}, \bibinfo{person}{Peter~A Boncz}, {and} \bibinfo{person}{Stavros Harizopoulos}.} \bibinfo{year}{2009}\natexlab{}.
\newblock \showarticletitle{Column-oriented database systems}.
\newblock \bibinfo{journal}{\emph{PVLDB}} \bibinfo{volume}{2}, \bibinfo{number}{2} (\bibinfo{year}{2009}), \bibinfo{pages}{1664--1665}.
\newblock


\bibitem[Abadi et~al\mbox{.}(2006)]%
        {abadi2006materialization}
\bibfield{author}{\bibinfo{person}{Daniel~J Abadi}, \bibinfo{person}{Daniel~S Myers}, \bibinfo{person}{David~J DeWitt}, {and} \bibinfo{person}{Samuel~R Madden}.} \bibinfo{year}{2006}\natexlab{}.
\newblock \showarticletitle{Materialization Strategies in a Column-Oriented DBMS}. In \bibinfo{booktitle}{\emph{ICDE}}. \bibinfo{pages}{466--475}.
\newblock


\bibitem[Armenatzoglou et~al\mbox{.}(2022)]%
        {armenatzoglou2022amazon}
\bibfield{author}{\bibinfo{person}{Nikos Armenatzoglou}, \bibinfo{person}{Sanuj Basu}, \bibinfo{person}{Naga Bhanoori}, \bibinfo{person}{Mengchu Cai}, \bibinfo{person}{Naresh Chainani}, \bibinfo{person}{Kiran Chinta}, \bibinfo{person}{Venkatraman Govindaraju}, \bibinfo{person}{Todd~J Green}, \bibinfo{person}{Monish Gupta}, \bibinfo{person}{Sebastian Hillig}, {et~al\mbox{.}}} \bibinfo{year}{2022}\natexlab{}.
\newblock \showarticletitle{Amazon Redshift re-invented}. In \bibinfo{booktitle}{\emph{SIGMOD}}. \bibinfo{pages}{2205--2217}.
\newblock


\bibitem[Beazley(2010)]%
        {beazley2010understanding}
\bibfield{author}{\bibinfo{person}{David Beazley}.} \bibinfo{year}{2010}\natexlab{}.
\newblock \showarticletitle{Understanding the Python {}}. In \bibinfo{booktitle}{\emph{PyCON Python Conference. Atlanta, Georgia}}. \bibinfo{pages}{1--62}.
\newblock


\bibitem[Brown et~al\mbox{.}(2020)]%
        {gpt3_2020}
\bibfield{author}{\bibinfo{person}{Tom Brown}, \bibinfo{person}{Benjamin Mann}, \bibinfo{person}{Nick Ryder}, \bibinfo{person}{Melanie Subbiah}, \bibinfo{person}{Jared~D Kaplan}, \bibinfo{person}{Prafulla Dhariwal}, \bibinfo{person}{Arvind Neelakantan}, \bibinfo{person}{Pranav Shyam}, \bibinfo{person}{Girish Sastry}, \bibinfo{person}{Amanda Askell}, {et~al\mbox{.}}} \bibinfo{year}{2020}\natexlab{}.
\newblock \showarticletitle{Language models are few-shot learners}.
\newblock \bibinfo{journal}{\emph{NIPS}}  \bibinfo{volume}{33} (\bibinfo{year}{2020}), \bibinfo{pages}{1877--1901}.
\newblock


\bibitem[Chao and Lee(1992)]%
        {chao1992estimating}
\bibfield{author}{\bibinfo{person}{Anne Chao} {and} \bibinfo{person}{Shen-Ming Lee}.} \bibinfo{year}{1992}\natexlab{}.
\newblock \showarticletitle{Estimating the number of classes via sample coverage}.
\newblock \bibinfo{journal}{\emph{Journal of the American statistical Association}} \bibinfo{volume}{87}, \bibinfo{number}{417} (\bibinfo{year}{1992}), \bibinfo{pages}{210--217}.
\newblock


\bibitem[Charikar et~al\mbox{.}(2000)]%
        {charikar2000towards}
\bibfield{author}{\bibinfo{person}{Moses Charikar}, \bibinfo{person}{Surajit Chaudhuri}, \bibinfo{person}{Rajeev Motwani}, {and} \bibinfo{person}{Vivek Narasayya}.} \bibinfo{year}{2000}\natexlab{}.
\newblock \showarticletitle{Towards estimation error guarantees for distinct values}. In \bibinfo{booktitle}{\emph{SIGMOD}}. \bibinfo{pages}{268--279}.
\newblock


\bibitem[Chen et~al\mbox{.}(2023)]%
        {chen2023workloadaware}
\bibfield{author}{\bibinfo{person}{Lixiang Chen}, \bibinfo{person}{Ruihao Chen}, \bibinfo{person}{Chengcheng Yang}, \bibinfo{person}{Yuxing Han}, \bibinfo{person}{Rong Zhang}, \bibinfo{person}{Xuan Zhou}, \bibinfo{person}{Peiquan Jin}, {and} \bibinfo{person}{Weining Qian}.} \bibinfo{year}{2023}\natexlab{}.
\newblock \showarticletitle{Workload-Aware Log-Structured Merge Key-Value Store for NVM-SSD Hybrid Storage}. In \bibinfo{booktitle}{\emph{ICDE}}. \bibinfo{pages}{2198--2210}.
\newblock


\bibitem[Chow and Liu(1968)]%
        {chow1968approximating}
\bibfield{author}{\bibinfo{person}{CKCN Chow} {and} \bibinfo{person}{Cong Liu}.} \bibinfo{year}{1968}\natexlab{}.
\newblock \showarticletitle{Approximating discrete probability distributions with dependence trees}.
\newblock \bibinfo{journal}{\emph{IEEE transactions on Information Theory}} \bibinfo{volume}{14}, \bibinfo{number}{3} (\bibinfo{year}{1968}), \bibinfo{pages}{462--467}.
\newblock


\bibitem[Cohen and Nezri(2019)]%
        {Cohen2019CardinalityEI}
\bibfield{author}{\bibinfo{person}{Reuven Cohen} {and} \bibinfo{person}{Yuval Nezri}.} \bibinfo{year}{2019}\natexlab{}.
\newblock \showarticletitle{Cardinality Estimation in a Virtualized Network Device Using Online Machine Learning}.
\newblock \bibinfo{journal}{\emph{IEEE/ACM Transactions on Networking}} \bibinfo{volume}{27}, \bibinfo{number}{5} (\bibinfo{year}{2019}), \bibinfo{pages}{2098--2110}.
\newblock


\bibitem[Dageville et~al\mbox{.}(2016)]%
        {dageville2016snowflake}
\bibfield{author}{\bibinfo{person}{Benoit Dageville}, \bibinfo{person}{Thierry Cruanes}, \bibinfo{person}{Marcin Zukowski}, \bibinfo{person}{Vadim Antonov}, \bibinfo{person}{Artin Avanes}, \bibinfo{person}{Jon Bock}, \bibinfo{person}{Jonathan Claybaugh}, \bibinfo{person}{Daniel Engovatov}, \bibinfo{person}{Martin Hentschel}, \bibinfo{person}{Jiansheng Huang}, {et~al\mbox{.}}} \bibinfo{year}{2016}\natexlab{}.
\newblock \showarticletitle{The snowflake elastic data warehouse}. In \bibinfo{booktitle}{\emph{SIGMOD}}. \bibinfo{pages}{215--226}.
\newblock


\bibitem[Dai et~al\mbox{.}(2020)]%
        {dai2020wisckey}
\bibfield{author}{\bibinfo{person}{Yifan Dai}, \bibinfo{person}{Yien Xu}, \bibinfo{person}{Aishwarya Ganesan}, \bibinfo{person}{Ramnatthan Alagappan}, \bibinfo{person}{Brian Kroth}, \bibinfo{person}{Andrea Arpaci-Dusseau}, {and} \bibinfo{person}{Remzi Arpaci-Dusseau}.} \bibinfo{year}{2020}\natexlab{}.
\newblock \showarticletitle{From $\{$WiscKey$\}$ to Bourbon: A Learned Index for $\{$Log-Structured$\}$ Merge Trees}. In \bibinfo{booktitle}{\emph{OSDI}}. \bibinfo{pages}{155--171}.
\newblock


\bibitem[Ding et~al\mbox{.}(2022)]%
        {ding2022sagedb}
\bibfield{author}{\bibinfo{person}{Jialin Ding}, \bibinfo{person}{Ryan Marcus}, \bibinfo{person}{Andreas Kipf}, \bibinfo{person}{Vikram Nathan}, \bibinfo{person}{Aniruddha Nrusimha}, \bibinfo{person}{Kapil Vaidya}, \bibinfo{person}{Alexander van Renen}, {and} \bibinfo{person}{Tim Kraska}.} \bibinfo{year}{2022}\natexlab{}.
\newblock \showarticletitle{SageDB: An Instance-Optimized Data Analytics System}.
\newblock \bibinfo{journal}{\emph{PVLDB}} \bibinfo{volume}{15}, \bibinfo{number}{13} (\bibinfo{year}{2022}), \bibinfo{pages}{4062--4078}.
\newblock


\bibitem[Do and Batzoglou(2008)]%
        {do2008expectation}
\bibfield{author}{\bibinfo{person}{Chuong~B Do} {and} \bibinfo{person}{Serafim Batzoglou}.} \bibinfo{year}{2008}\natexlab{}.
\newblock \showarticletitle{What is the expectation maximization algorithm?}
\newblock \bibinfo{journal}{\emph{Nature biotechnology}} \bibinfo{volume}{26}, \bibinfo{number}{8} (\bibinfo{year}{2008}), \bibinfo{pages}{897--899}.
\newblock


\bibitem[Dutt et~al\mbox{.}(2019)]%
        {dutt2019selectivity}
\bibfield{author}{\bibinfo{person}{Anshuman Dutt}, \bibinfo{person}{Chi Wang}, \bibinfo{person}{Azade Nazi}, \bibinfo{person}{Srikanth Kandula}, \bibinfo{person}{Vivek Narasayya}, {and} \bibinfo{person}{Surajit Chaudhuri}.} \bibinfo{year}{2019}\natexlab{}.
\newblock \showarticletitle{Selectivity estimation for range predicates using lightweight models}.
\newblock \bibinfo{journal}{\emph{PVLDB}} \bibinfo{volume}{12}, \bibinfo{number}{9} (\bibinfo{year}{2019}), \bibinfo{pages}{1044--1057}.
\newblock


\bibitem[Flajolet et~al\mbox{.}(2007)]%
        {flajolet2007hyperloglog}
\bibfield{author}{\bibinfo{person}{Philippe Flajolet}, \bibinfo{person}{{\'E}ric Fusy}, \bibinfo{person}{Olivier Gandouet}, {and} \bibinfo{person}{Fr{\'e}d{\'e}ric Meunier}.} \bibinfo{year}{2007}\natexlab{}.
\newblock \showarticletitle{Hyperloglog: the analysis of a near-optimal cardinality estimation algorithm}. In \bibinfo{booktitle}{\emph{Discrete Mathematics and Theoretical Computer Science}}. \bibinfo{pages}{137--156}.
\newblock


\bibitem[Gonzales et~al\mbox{.}(2017)]%
        {agrum2017}
\bibfield{author}{\bibinfo{person}{Christophe Gonzales}, \bibinfo{person}{Lionel Torti}, {and} \bibinfo{person}{Pierre-Henri Wuillemin}.} \bibinfo{year}{2017}\natexlab{}.
\newblock \showarticletitle{{aGrUM: a Graphical Universal Model framework}}. In \bibinfo{booktitle}{\emph{IEA/AIE}}. \bibinfo{pages}{171--177}.
\newblock


\bibitem[Graefe(1995)]%
        {graefe1995cascades}
\bibfield{author}{\bibinfo{person}{Goetz Graefe}.} \bibinfo{year}{1995}\natexlab{}.
\newblock \showarticletitle{The cascades framework for query optimization}.
\newblock \bibinfo{journal}{\emph{IEEE Data Eng. Bull.}} \bibinfo{volume}{18}, \bibinfo{number}{3} (\bibinfo{year}{1995}), \bibinfo{pages}{19--29}.
\newblock


\bibitem[Han et~al\mbox{.}(2021a)]%
        {han2021autonomous}
\bibfield{author}{\bibinfo{person}{Yue Han}, \bibinfo{person}{Guoliang Li}, \bibinfo{person}{Haitao Yuan}, {and} \bibinfo{person}{Ji Sun}.} \bibinfo{year}{2021}\natexlab{a}.
\newblock \showarticletitle{An autonomous materialized view management system with deep reinforcement learning}. In \bibinfo{booktitle}{\emph{ICDE}}. \bibinfo{pages}{2159--2164}.
\newblock


\bibitem[Han et~al\mbox{.}(2021b)]%
        {cardestbench}
\bibfield{author}{\bibinfo{person}{Yuxing Han}, \bibinfo{person}{Ziniu Wu}, \bibinfo{person}{Peizhi Wu}, \bibinfo{person}{Rong Zhu}, \bibinfo{person}{Jingyi Yang}, \bibinfo{person}{Liang~Wei Tan}, \bibinfo{person}{Kai Zeng}, \bibinfo{person}{Gao Cong}, \bibinfo{person}{Yanzhao Qin}, \bibinfo{person}{Andreas Pfadler}, \bibinfo{person}{Zhengping Qian}, \bibinfo{person}{Jingren Zhou}, \bibinfo{person}{Jiangneng Li}, {and} \bibinfo{person}{Bin Cui}.} \bibinfo{year}{2021}\natexlab{b}.
\newblock \showarticletitle{Cardinality Estimation in DBMS: A Comprehensive Benchmark Evaluation}.
\newblock \bibinfo{journal}{\emph{PVLDB}} \bibinfo{volume}{15}, \bibinfo{number}{4} (\bibinfo{year}{2021}), \bibinfo{pages}{752--765}.
\newblock


\bibitem[Heule et~al\mbox{.}(2013)]%
        {heule2013hyperloglog}
\bibfield{author}{\bibinfo{person}{Stefan Heule}, \bibinfo{person}{Marc Nunkesser}, {and} \bibinfo{person}{Alexander Hall}.} \bibinfo{year}{2013}\natexlab{}.
\newblock \showarticletitle{Hyperloglog in practice: Algorithmic engineering of a state of the art cardinality estimation algorithm}. In \bibinfo{booktitle}{\emph{{EDBT/ICDT}}}. \bibinfo{pages}{683--692}.
\newblock


\bibitem[Hilprecht and Binnig(2021)]%
        {hilprecht2021one}
\bibfield{author}{\bibinfo{person}{Benjamin Hilprecht} {and} \bibinfo{person}{Carsten Binnig}.} \bibinfo{year}{2021}\natexlab{}.
\newblock \showarticletitle{One model to rule them all: towards zero-shot learning for databases}.
\newblock \bibinfo{journal}{\emph{arXiv:2105.00642}} (\bibinfo{year}{2021}).
\newblock


\bibitem[Hilprecht and Binnig(2022)]%
        {hizeroshot2022}
\bibfield{author}{\bibinfo{person}{Benjamin Hilprecht} {and} \bibinfo{person}{Carsten Binnig}.} \bibinfo{year}{2022}\natexlab{}.
\newblock \showarticletitle{Zero-Shot Cost Models for out-of-the-Box Learned Cost Prediction}.
\newblock \bibinfo{journal}{\emph{PVLDB}} \bibinfo{volume}{15}, \bibinfo{number}{11} (\bibinfo{year}{2022}), \bibinfo{pages}{2361–2374}.
\newblock


\bibitem[Hilprecht et~al\mbox{.}(2020)]%
        {hilprecht2019deepdb}
\bibfield{author}{\bibinfo{person}{Benjamin Hilprecht}, \bibinfo{person}{Andreas Schmidt}, \bibinfo{person}{Moritz Kulessa}, \bibinfo{person}{Alejandro Molina}, \bibinfo{person}{Kristian Kersting}, {and} \bibinfo{person}{Carsten Binnig}.} \bibinfo{year}{2020}\natexlab{}.
\newblock \showarticletitle{DeepDB: learn from data, not from queries!}
\newblock \bibinfo{journal}{\emph{PVLDB}} \bibinfo{volume}{13}, \bibinfo{number}{7} (\bibinfo{year}{2020}), \bibinfo{pages}{992--1005}.
\newblock


\bibitem[Huang et~al\mbox{.}(2020)]%
        {HuangLCFMXSTZHW20}
\bibfield{author}{\bibinfo{person}{Dongxu Huang}, \bibinfo{person}{Qi Liu}, \bibinfo{person}{Qiu Cui}, \bibinfo{person}{Zhuhe Fang}, \bibinfo{person}{Xiaoyu Ma}, \bibinfo{person}{Fei Xu}, \bibinfo{person}{Li Shen}, \bibinfo{person}{Liu Tang}, \bibinfo{person}{Yuxing Zhou}, \bibinfo{person}{Menglong Huang}, \bibinfo{person}{Wan Wei}, \bibinfo{person}{Cong Liu}, \bibinfo{person}{Jian Zhang}, \bibinfo{person}{Jianjun Li}, \bibinfo{person}{Xuelian Wu}, \bibinfo{person}{Lingyu Song}, \bibinfo{person}{Ruoxi Sun}, \bibinfo{person}{Shuaipeng Yu}, \bibinfo{person}{Lei Zhao}, \bibinfo{person}{Nicholas Cameron}, \bibinfo{person}{Liquan Pei}, {and} \bibinfo{person}{Xin Tang}.} \bibinfo{year}{2020}\natexlab{}.
\newblock \showarticletitle{TiDB: {A} Raft-based {HTAP} Database}.
\newblock \bibinfo{journal}{\emph{PVLDB}} \bibinfo{volume}{13}, \bibinfo{number}{12} (\bibinfo{year}{2020}), \bibinfo{pages}{3072--3084}.
\newblock


\bibitem[Ives and Taylor(2008)]%
        {ives2008sideways}
\bibfield{author}{\bibinfo{person}{Zachary~G Ives} {and} \bibinfo{person}{Nicholas~E Taylor}.} \bibinfo{year}{2008}\natexlab{}.
\newblock \showarticletitle{Sideways information passing for push-style query processing}. In \bibinfo{booktitle}{\emph{ICDE}}. \bibinfo{pages}{774--783}.
\newblock


\bibitem[Kipf et~al\mbox{.}(2019)]%
        {MSCN}
\bibfield{author}{\bibinfo{person}{Andreas Kipf}, \bibinfo{person}{Thomas Kipf}, \bibinfo{person}{Bernhard Radke}, \bibinfo{person}{Viktor Leis}, \bibinfo{person}{Peter Boncz}, {and} \bibinfo{person}{Alfons Kemper}.} \bibinfo{year}{2019}\natexlab{}.
\newblock \showarticletitle{Learned cardinalities: Estimating correlated joins with deep learning}. In \bibinfo{booktitle}{\emph{CIDR}}.
\newblock


\bibitem[Koller and Friedman(2009)]%
        {koller2009probabilistic}
\bibfield{author}{\bibinfo{person}{Daphne Koller} {and} \bibinfo{person}{Nir Friedman}.} \bibinfo{year}{2009}\natexlab{}.
\newblock \bibinfo{booktitle}{\emph{Probabilistic graphical models: principles and techniques}}.
\newblock \bibinfo{publisher}{MIT press}.
\newblock


\bibitem[Lamb et~al\mbox{.}(2012)]%
        {lamb2012vertica}
\bibfield{author}{\bibinfo{person}{Andrew Lamb}, \bibinfo{person}{Matt Fuller}, \bibinfo{person}{Ramakrishna Varadarajan}, \bibinfo{person}{Nga Tran}, \bibinfo{person}{Ben Vandiver}, \bibinfo{person}{Lyric Doshi}, {and} \bibinfo{person}{Chuck Bear}.} \bibinfo{year}{2012}\natexlab{}.
\newblock \showarticletitle{The Vertica Analytic Database: C-Store 7 Years Later}.
\newblock \bibinfo{journal}{\emph{PVLDB}} \bibinfo{volume}{5}, \bibinfo{number}{12} (\bibinfo{year}{2012}), \bibinfo{pages}{1790--1801}.
\newblock


\bibitem[Lee et~al\mbox{.}(2023)]%
        {lee2023analyzing}
\bibfield{author}{\bibinfo{person}{Kukjin Lee}, \bibinfo{person}{Anshuman Dutt}, \bibinfo{person}{Vivek Narasayya}, {and} \bibinfo{person}{Surajit Chaudhuri}.} \bibinfo{year}{2023}\natexlab{}.
\newblock \showarticletitle{Analyzing the Impact of Cardinality Estimation on Execution Plans in Microsoft SQL Server}.
\newblock \bibinfo{journal}{\emph{PVLDB}} \bibinfo{volume}{16}, \bibinfo{number}{11} (\bibinfo{year}{2023}), \bibinfo{pages}{2871--2883}.
\newblock


\bibitem[Leis et~al\mbox{.}(2015)]%
        {leis2015good}
\bibfield{author}{\bibinfo{person}{Viktor Leis}, \bibinfo{person}{Andrey Gubichev}, \bibinfo{person}{Atanas Mirchev}, \bibinfo{person}{Peter Boncz}, \bibinfo{person}{Alfons Kemper}, {and} \bibinfo{person}{Thomas Neumann}.} \bibinfo{year}{2015}\natexlab{}.
\newblock \showarticletitle{How good are query optimizers, really?}
\newblock \bibinfo{journal}{\emph{PVLDB}} \bibinfo{volume}{9}, \bibinfo{number}{3} (\bibinfo{year}{2015}), \bibinfo{pages}{204--215}.
\newblock


\bibitem[Leis et~al\mbox{.}(2018)]%
        {leis2018query}
\bibfield{author}{\bibinfo{person}{Viktor Leis}, \bibinfo{person}{Bernhard Radke}, \bibinfo{person}{Andrey Gubichev}, \bibinfo{person}{Atanas Mirchev}, \bibinfo{person}{Peter Boncz}, \bibinfo{person}{Alfons Kemper}, {and} \bibinfo{person}{Thomas Neumann}.} \bibinfo{year}{2018}\natexlab{}.
\newblock \showarticletitle{Query optimization through the looking glass, and what we found running the join order benchmark}.
\newblock \bibinfo{journal}{\emph{PVLDB}} \bibinfo{volume}{27}, \bibinfo{number}{5} (\bibinfo{year}{2018}), \bibinfo{pages}{643--668}.
\newblock


\bibitem[Li et~al\mbox{.}(2021)]%
        {opengauss}
\bibfield{author}{\bibinfo{person}{Guoliang Li}, \bibinfo{person}{Xuanhe Zhou}, \bibinfo{person}{Ji Sun}, \bibinfo{person}{Xiang Yu}, \bibinfo{person}{Yue Han}, \bibinfo{person}{Lianyuan Jin}, \bibinfo{person}{Wenbo Li}, \bibinfo{person}{Tianqing Wang}, {and} \bibinfo{person}{Shifu Li}.} \bibinfo{year}{2021}\natexlab{}.
\newblock \showarticletitle{opengauss: An autonomous database system}.
\newblock \bibinfo{journal}{\emph{PVLDB}} \bibinfo{volume}{14}, \bibinfo{number}{12} (\bibinfo{year}{2021}), \bibinfo{pages}{3028--3042}.
\newblock


\bibitem[Liang et~al\mbox{.}(2019)]%
        {liang2019opportunistic}
\bibfield{author}{\bibinfo{person}{Xi Liang}, \bibinfo{person}{Aaron~J Elmore}, {and} \bibinfo{person}{Sanjay Krishnan}.} \bibinfo{year}{2019}\natexlab{}.
\newblock \showarticletitle{Opportunistic view materialization with deep reinforcement learning}.
\newblock \bibinfo{journal}{\emph{arXiv:1903.01363}} (\bibinfo{year}{2019}).
\newblock


\bibitem[Liu et~al\mbox{.}(2021)]%
        {liu2021fauce}
\bibfield{author}{\bibinfo{person}{Jie Liu}, \bibinfo{person}{Wenqian Dong}, \bibinfo{person}{Qingqing Zhou}, {and} \bibinfo{person}{Dong Li}.} \bibinfo{year}{2021}\natexlab{}.
\newblock \showarticletitle{Fauce: fast and accurate deep ensembles with uncertainty for cardinality estimation}.
\newblock \bibinfo{journal}{\emph{PVLDB}} \bibinfo{volume}{14}, \bibinfo{number}{11} (\bibinfo{year}{2021}), \bibinfo{pages}{1950--1963}.
\newblock


\bibitem[Loeliger(2004)]%
        {loeliger2004introduction}
\bibfield{author}{\bibinfo{person}{H-A Loeliger}.} \bibinfo{year}{2004}\natexlab{}.
\newblock \showarticletitle{An introduction to factor graphs}.
\newblock \bibinfo{journal}{\emph{IEEE Signal Processing Magazine}} \bibinfo{volume}{21}, \bibinfo{number}{1} (\bibinfo{year}{2004}), \bibinfo{pages}{28--41}.
\newblock


\bibitem[Negi et~al\mbox{.}(2023)]%
        {negi2023robust}
\bibfield{author}{\bibinfo{person}{Parimarjan Negi}, \bibinfo{person}{Ziniu Wu}, \bibinfo{person}{Andreas Kipf}, \bibinfo{person}{Nesime Tatbul}, \bibinfo{person}{Ryan Marcus}, \bibinfo{person}{Sam Madden}, \bibinfo{person}{Tim Kraska}, {and} \bibinfo{person}{Mohammad Alizadeh}.} \bibinfo{year}{2023}\natexlab{}.
\newblock \showarticletitle{Robust Query Driven Cardinality Estimation under Changing Workloads}.
\newblock \bibinfo{journal}{\emph{PVLDB}} \bibinfo{volume}{16}, \bibinfo{number}{6} (\bibinfo{year}{2023}), \bibinfo{pages}{1520--1533}.
\newblock


\bibitem[Sarkar et~al\mbox{.}(2021)]%
        {lsmanalysis}
\bibfield{author}{\bibinfo{person}{Subhadeep Sarkar}, \bibinfo{person}{Dimitris Staratzis}, \bibinfo{person}{Ziehen Zhu}, {and} \bibinfo{person}{Manos Athanassoulis}.} \bibinfo{year}{2021}\natexlab{}.
\newblock \showarticletitle{Constructing and Analyzing the LSM Compaction Design Space}.
\newblock \bibinfo{journal}{\emph{PVLDB}} \bibinfo{volume}{14}, \bibinfo{number}{11} (\bibinfo{year}{2021}), \bibinfo{pages}{2216--2229}.
\newblock


\bibitem[Saxena et~al\mbox{.}(2023)]%
        {saxena2023auto}
\bibfield{author}{\bibinfo{person}{Gaurav Saxena}, \bibinfo{person}{Mohammad Rahman}, \bibinfo{person}{Naresh Chainani}, \bibinfo{person}{Chunbin Lin}, \bibinfo{person}{George Caragea}, \bibinfo{person}{Fahim Chowdhury}, \bibinfo{person}{Ryan Marcus}, \bibinfo{person}{Tim Kraska}, \bibinfo{person}{Ippokratis Pandis}, {and} \bibinfo{person}{Balakrishnan Narayanaswamy}.} \bibinfo{year}{2023}\natexlab{}.
\newblock \showarticletitle{Auto-WLM: Machine learning enhanced workload management in Amazon Redshift}. In \bibinfo{booktitle}{\emph{Companion of the International Conference on Management of Data, {SIGMOD/PODS}}}. \bibinfo{pages}{225--237}.
\newblock


\bibitem[Schad et~al\mbox{.}(2010)]%
        {schad2010runtime}
\bibfield{author}{\bibinfo{person}{J{\"o}rg Schad}, \bibinfo{person}{Jens Dittrich}, {and} \bibinfo{person}{Jorge-Arnulfo Quian{\'e}-Ruiz}.} \bibinfo{year}{2010}\natexlab{}.
\newblock \showarticletitle{Runtime measurements in the cloud: observing, analyzing, and reducing variance}.
\newblock \bibinfo{journal}{\emph{PVLDB}} \bibinfo{volume}{3}, \bibinfo{number}{1-2} (\bibinfo{year}{2010}), \bibinfo{pages}{460--471}.
\newblock


\bibitem[Selinger et~al\mbox{.}(1979)]%
        {selinger1979access}
\bibfield{author}{\bibinfo{person}{P~Griffiths Selinger}, \bibinfo{person}{Morton~M Astrahan}, \bibinfo{person}{Donald~D Chamberlin}, \bibinfo{person}{Raymond~A Lorie}, {and} \bibinfo{person}{Thomas~G Price}.} \bibinfo{year}{1979}\natexlab{}.
\newblock \showarticletitle{Access path selection in a relational database management system}. In \bibinfo{booktitle}{\emph{SIGMOD}}. \bibinfo{pages}{23--34}.
\newblock


\bibitem[Seshadri et~al\mbox{.}(1996)]%
        {seshadri1996cost}
\bibfield{author}{\bibinfo{person}{Praveen Seshadri}, \bibinfo{person}{Joseph~M Hellerstein}, \bibinfo{person}{Hamid Pirahesh}, \bibinfo{person}{TY~Cliff Leung}, \bibinfo{person}{Raghu Ramakrishnan}, \bibinfo{person}{Divesh Srivastava}, \bibinfo{person}{Peter~J Stuckey}, {and} \bibinfo{person}{S Sudarshan}.} \bibinfo{year}{1996}\natexlab{}.
\newblock \showarticletitle{Cost-based optimization for magic: Algebra and implementation}. In \bibinfo{booktitle}{\emph{SIGMOD}}. \bibinfo{pages}{435--446}.
\newblock


\bibitem[Shen et~al\mbox{.}(2023)]%
        {shen2023rover}
\bibfield{author}{\bibinfo{person}{Yu Shen}, \bibinfo{person}{Xinyuyang Ren}, \bibinfo{person}{Yupeng Lu}, \bibinfo{person}{Huaijun Jiang}, \bibinfo{person}{Huanyong Xu}, \bibinfo{person}{Di Peng}, \bibinfo{person}{Yang Li}, \bibinfo{person}{Wentao Zhang}, {and} \bibinfo{person}{Bin Cui}.} \bibinfo{year}{2023}\natexlab{}.
\newblock \showarticletitle{Rover: An online Spark SQL tuning service via generalized transfer learning}. In \bibinfo{booktitle}{\emph{SIGKDD}}. \bibinfo{pages}{4800--4812}.
\newblock


\bibitem[Shrinivas et~al\mbox{.}(2013)]%
        {shrinivas2013materialization}
\bibfield{author}{\bibinfo{person}{Lakshmikant Shrinivas}, \bibinfo{person}{Sreenath Bodagala}, \bibinfo{person}{Ramakrishna Varadarajan}, \bibinfo{person}{Ariel Cary}, \bibinfo{person}{Vivek Bharathan}, {and} \bibinfo{person}{Chuck Bear}.} \bibinfo{year}{2013}\natexlab{}.
\newblock \showarticletitle{Materialization Strategies in the Vertica Analytic Database: Lessons Learned}. In \bibinfo{booktitle}{\emph{ICDE}}. \bibinfo{pages}{1196--1207}.
\newblock


\bibitem[Siddiqui et~al\mbox{.}(2020)]%
        {siddiqui2020cost}
\bibfield{author}{\bibinfo{person}{Tarique Siddiqui}, \bibinfo{person}{Alekh Jindal}, \bibinfo{person}{Shi Qiao}, \bibinfo{person}{Hiren Patel}, {and} \bibinfo{person}{Wangchao Le}.} \bibinfo{year}{2020}\natexlab{}.
\newblock \showarticletitle{Cost models for big data query processing: Learning, retrofitting, and our findings}. In \bibinfo{booktitle}{\emph{SIGMOD}}. \bibinfo{pages}{99--113}.
\newblock


\bibitem[Sun and Li(2019)]%
        {suncostmodels2019}
\bibfield{author}{\bibinfo{person}{Ji Sun} {and} \bibinfo{person}{Guoliang Li}.} \bibinfo{year}{2019}\natexlab{}.
\newblock \showarticletitle{An End-to-End Learning-Based Cost Estimator}.
\newblock \bibinfo{journal}{\emph{PVLDB}} \bibinfo{volume}{13}, \bibinfo{number}{3} (\bibinfo{year}{2019}), \bibinfo{pages}{307--319}.
\newblock


\bibitem[Sun et~al\mbox{.}(2021)]%
        {sun2021learned}
\bibfield{author}{\bibinfo{person}{Ji Sun}, \bibinfo{person}{Jintao Zhang}, \bibinfo{person}{Zhaoyan Sun}, \bibinfo{person}{Guoliang Li}, {and} \bibinfo{person}{Nan Tang}.} \bibinfo{year}{2021}\natexlab{}.
\newblock \showarticletitle{Learned cardinality estimation: A design space exploration and a comparative evaluation}.
\newblock \bibinfo{journal}{\emph{PVLDB}} \bibinfo{volume}{15}, \bibinfo{number}{1} (\bibinfo{year}{2021}), \bibinfo{pages}{85--97}.
\newblock


\bibitem[Sun et~al\mbox{.}(2023)]%
        {sun2023presto}
\bibfield{author}{\bibinfo{person}{Yutian Sun}, \bibinfo{person}{Tim Meehan}, \bibinfo{person}{Rebecca Schlussel}, \bibinfo{person}{Wenlei Xie}, \bibinfo{person}{Masha Basmanova}, \bibinfo{person}{Orri Erling}, \bibinfo{person}{Andrii Rosa}, \bibinfo{person}{Shixuan Fan}, \bibinfo{person}{Rongrong Zhong}, \bibinfo{person}{Arun Thirupathi}, {et~al\mbox{.}}} \bibinfo{year}{2023}\natexlab{}.
\newblock \showarticletitle{Presto: A Decade of SQL Analytics at Meta}.
\newblock \bibinfo{journal}{\emph{SIGMOD}} \bibinfo{volume}{1}, \bibinfo{number}{2} (\bibinfo{year}{2023}), \bibinfo{pages}{1--25}.
\newblock


\bibitem[Tang et~al\mbox{.}(2020)]%
        {tang2020xindex}
\bibfield{author}{\bibinfo{person}{Chuzhe Tang}, \bibinfo{person}{Youyun Wang}, \bibinfo{person}{Zhiyuan Dong}, \bibinfo{person}{Gansen Hu}, \bibinfo{person}{Zhaoguo Wang}, \bibinfo{person}{Minjie Wang}, {and} \bibinfo{person}{Haibo Chen}.} \bibinfo{year}{2020}\natexlab{}.
\newblock \showarticletitle{XIndex: a scalable learned index for multicore data storage}. In \bibinfo{booktitle}{\emph{PPoPP}}. \bibinfo{pages}{308--320}.
\newblock


\bibitem[Thirumuruganathan et~al\mbox{.}(2022)]%
        {thirumuruganathan2022prediction}
\bibfield{author}{\bibinfo{person}{Saravanan Thirumuruganathan}, \bibinfo{person}{Suraj Shetiya}, \bibinfo{person}{Nick Koudas}, {and} \bibinfo{person}{Gautam Das}.} \bibinfo{year}{2022}\natexlab{}.
\newblock \showarticletitle{Prediction Intervals for Learned Cardinality Estimation: An Experimental Evaluation}. In \bibinfo{booktitle}{\emph{ICDE}}. \bibinfo{pages}{3051--3064}.
\newblock


\bibitem[Van~Aken et~al\mbox{.}(2017)]%
        {van2017automatic}
\bibfield{author}{\bibinfo{person}{Dana Van~Aken}, \bibinfo{person}{Andrew Pavlo}, \bibinfo{person}{Geoffrey~J Gordon}, {and} \bibinfo{person}{Bohan Zhang}.} \bibinfo{year}{2017}\natexlab{}.
\newblock \showarticletitle{Automatic database management system tuning through large-scale machine learning}. In \bibinfo{booktitle}{\emph{SIGMOD}}. \bibinfo{pages}{1009--1024}.
\newblock


\bibitem[Wang et~al\mbox{.}(2021)]%
        {wang2021face}
\bibfield{author}{\bibinfo{person}{Jiayi Wang}, \bibinfo{person}{Chengliang Chai}, \bibinfo{person}{Jiabin Liu}, {and} \bibinfo{person}{Guoliang Li}.} \bibinfo{year}{2021}\natexlab{}.
\newblock \showarticletitle{FACE: A normalizing flow based cardinality estimator}.
\newblock \bibinfo{journal}{\emph{PVLDB}} \bibinfo{volume}{15}, \bibinfo{number}{1} (\bibinfo{year}{2021}), \bibinfo{pages}{72--84}.
\newblock


\bibitem[Wei et~al\mbox{.}(2022)]%
        {wei2022emergent}
\bibfield{author}{\bibinfo{person}{Jason Wei}, \bibinfo{person}{Yi Tay}, \bibinfo{person}{Rishi Bommasani}, \bibinfo{person}{Colin Raffel}, \bibinfo{person}{Barret Zoph}, \bibinfo{person}{Sebastian Borgeaud}, \bibinfo{person}{Dani Yogatama}, \bibinfo{person}{Maarten Bosma}, \bibinfo{person}{Denny Zhou}, \bibinfo{person}{Donald Metzler}, {et~al\mbox{.}}} \bibinfo{year}{2022}\natexlab{}.
\newblock \showarticletitle{Emergent abilities of large language models}.
\newblock \bibinfo{journal}{\emph{arXiv:2206.07682}} (\bibinfo{year}{2022}).
\newblock


\bibitem[Wu et~al\mbox{.}(2021)]%
        {ndvwu2021}
\bibfield{author}{\bibinfo{person}{Renzhi Wu}, \bibinfo{person}{Bolin Ding}, \bibinfo{person}{Xu Chu}, \bibinfo{person}{Zhewei Wei}, \bibinfo{person}{Xiening Dai}, \bibinfo{person}{Tao Guan}, {and} \bibinfo{person}{Jingren Zhou}.} \bibinfo{year}{2021}\natexlab{}.
\newblock \showarticletitle{Learning to Be a Statistician: Learned Estimator for Number of Distinct Values}.
\newblock \bibinfo{journal}{\emph{PVLDB}} \bibinfo{volume}{15}, \bibinfo{number}{2} (\bibinfo{year}{2021}), \bibinfo{pages}{272--284}.
\newblock


\bibitem[Wu et~al\mbox{.}(2023)]%
        {wu2023factorjoin}
\bibfield{author}{\bibinfo{person}{Ziniu Wu}, \bibinfo{person}{Parimarjan Negi}, \bibinfo{person}{Mohammad Alizadeh}, \bibinfo{person}{Tim Kraska}, {and} \bibinfo{person}{Samuel Madden}.} \bibinfo{year}{2023}\natexlab{}.
\newblock \showarticletitle{FactorJoin: A New Cardinality Estimation Framework for Join Queries}.
\newblock \bibinfo{journal}{\emph{SIGMOD}} \bibinfo{volume}{1}, \bibinfo{number}{1} (\bibinfo{year}{2023}), \bibinfo{pages}{1--27}.
\newblock


\bibitem[Wu and Shaikhha(2020)]%
        {wu2020bayescard}
\bibfield{author}{\bibinfo{person}{Ziniu Wu} {and} \bibinfo{person}{Amir Shaikhha}.} \bibinfo{year}{2020}\natexlab{}.
\newblock \showarticletitle{BayesCard: A Unified Bayesian Framework for Cardinality Estimation}.
\newblock \bibinfo{journal}{\emph{arXiv:2012.14743}} (\bibinfo{year}{2020}).
\newblock


\bibitem[Wu et~al\mbox{.}(2022)]%
        {wu2021unified}
\bibfield{author}{\bibinfo{person}{Ziniu Wu}, \bibinfo{person}{Peilun Yang}, \bibinfo{person}{Pei Yu}, \bibinfo{person}{Rong Zhu}, \bibinfo{person}{Yuxing Han}, \bibinfo{person}{Yaliang Li}, \bibinfo{person}{Defu Lian}, \bibinfo{person}{Kai Zeng}, {and} \bibinfo{person}{Jingren Zhou}.} \bibinfo{year}{2022}\natexlab{}.
\newblock \showarticletitle{A Unified Transferable Model for ML-Enhanced DBMS}.
\newblock \bibinfo{journal}{\emph{CIDR}} (\bibinfo{year}{2022}).
\newblock


\bibitem[Yang et~al\mbox{.}(2021)]%
        {yang2020neurocard}
\bibfield{author}{\bibinfo{person}{Zongheng Yang}, \bibinfo{person}{Amog Kamsetty}, \bibinfo{person}{Sifei Luan}, \bibinfo{person}{Eric Liang}, \bibinfo{person}{Yan Duan}, \bibinfo{person}{Xi Chen}, {and} \bibinfo{person}{Ion Stoica}.} \bibinfo{year}{2021}\natexlab{}.
\newblock \showarticletitle{NeuroCard: One Cardinality Estimator for All Tables}.
\newblock \bibinfo{journal}{\emph{PVLDB}} \bibinfo{volume}{14}, \bibinfo{number}{1} (\bibinfo{year}{2021}), \bibinfo{pages}{61--73}.
\newblock


\bibitem[Yang et~al\mbox{.}(2019)]%
        {yang2019deep}
\bibfield{author}{\bibinfo{person}{Zongheng Yang}, \bibinfo{person}{Eric Liang}, \bibinfo{person}{Amog Kamsetty}, \bibinfo{person}{Chenggang Wu}, \bibinfo{person}{Yan Duan}, \bibinfo{person}{Xi Chen}, \bibinfo{person}{Pieter Abbeel}, \bibinfo{person}{Joseph~M Hellerstein}, \bibinfo{person}{Sanjay Krishnan}, {and} \bibinfo{person}{Ion Stoica}.} \bibinfo{year}{2019}\natexlab{}.
\newblock \showarticletitle{Deep unsupervised cardinality estimation}.
\newblock \bibinfo{journal}{\emph{PVLDB}} \bibinfo{volume}{13}, \bibinfo{number}{3} (\bibinfo{year}{2019}), \bibinfo{pages}{279--292}.
\newblock


\bibitem[Zhan et~al\mbox{.}(2019)]%
        {zhan2019analyticdb}
\bibfield{author}{\bibinfo{person}{Chaoqun Zhan}, \bibinfo{person}{Maomeng Su}, \bibinfo{person}{Chuangxian Wei}, \bibinfo{person}{Xiaoqiang Peng}, \bibinfo{person}{Liang Lin}, \bibinfo{person}{Sheng Wang}, \bibinfo{person}{Zhe Chen}, \bibinfo{person}{Feifei Li}, \bibinfo{person}{Yue Pan}, \bibinfo{person}{Fang Zheng}, {et~al\mbox{.}}} \bibinfo{year}{2019}\natexlab{}.
\newblock \showarticletitle{AnalyticDB: real-time OLAP database system at Alibaba cloud}.
\newblock \bibinfo{journal}{\emph{PVLDB}} \bibinfo{volume}{12}, \bibinfo{number}{12} (\bibinfo{year}{2019}), \bibinfo{pages}{2059--2070}.
\newblock


\bibitem[Zhang et~al\mbox{.}(2019)]%
        {zhang2019end}
\bibfield{author}{\bibinfo{person}{Ji Zhang}, \bibinfo{person}{Yu Liu}, \bibinfo{person}{Ke Zhou}, \bibinfo{person}{Guoliang Li}, \bibinfo{person}{Zhili Xiao}, \bibinfo{person}{Bin Cheng}, \bibinfo{person}{Jiashu Xing}, \bibinfo{person}{Yangtao Wang}, \bibinfo{person}{Tianheng Cheng}, \bibinfo{person}{Li Liu}, {et~al\mbox{.}}} \bibinfo{year}{2019}\natexlab{}.
\newblock \showarticletitle{An end-to-end automatic cloud database tuning system using deep reinforcement learning}. In \bibinfo{booktitle}{\emph{SIGMOD}}. \bibinfo{pages}{415--432}.
\newblock


\bibitem[Zhang et~al\mbox{.}(2021)]%
        {ZhangWCJT0Z021}
\bibfield{author}{\bibinfo{person}{Xinyi Zhang}, \bibinfo{person}{Hong Wu}, \bibinfo{person}{Zhuo Chang}, \bibinfo{person}{Shuowei Jin}, \bibinfo{person}{Jian Tan}, \bibinfo{person}{Feifei Li}, \bibinfo{person}{Tieying Zhang}, {and} \bibinfo{person}{Bin Cui}.} \bibinfo{year}{2021}\natexlab{}.
\newblock \showarticletitle{ResTune: Resource Oriented Tuning Boosted by Meta-Learning for Cloud Databases}. In \bibinfo{booktitle}{\emph{{SIGMOD}}}. \bibinfo{pages}{2102--2114}.
\newblock


\bibitem[Zhu et~al\mbox{.}(2021)]%
        {zhu2020flat}
\bibfield{author}{\bibinfo{person}{Rong Zhu}, \bibinfo{person}{Ziniu Wu}, \bibinfo{person}{Yuxing Han}, \bibinfo{person}{Kai Zeng}, \bibinfo{person}{Andreas Pfadler}, \bibinfo{person}{Zhengping Qian}, \bibinfo{person}{Jingren Zhou}, {and} \bibinfo{person}{Bin Cui}.} \bibinfo{year}{2021}\natexlab{}.
\newblock \showarticletitle{FLAT: Fast, Lightweight and Accurate Method for Cardinality Estimation}.
\newblock \bibinfo{journal}{\emph{PVLDB}} \bibinfo{volume}{14}, \bibinfo{number}{9} (\bibinfo{year}{2021}), \bibinfo{pages}{1489--1502}.
\newblock


\end{thebibliography}
